\documentclass{JINST}

\newcommand{\ops}{o-Ps}
\newcommand{\mops}{o-Ps'}
\newcommand{\pps}{p-Ps}

\newcommand{\invdecay}{$\rm{o-Ps}\to \rm{invisible}$}
\newcommand{\binvdecay}{Br(\invdecay)}

\title{Positronium Portal into Hidden Sector:\\
A new Experiment to Search for Mirror Dark Matter}

\author{Paolo Crivelli$^a$, Alexander Belov $^b$, Ulisse Gendotti$^c$, Sergei Gninenko$^b$, Andr\' e~Rubbia$^c$\\
\llap{$^a$}Instituto de Fisica, UFRJ,\\
 Rio de Janeiro, Brazil\\
\llap{$^b$}Institute for Nuclear Research (INR),\\
  Moscow, Russia\\
\llap{$^c$}ETH Zurich,\\
 101 Raemistrasse 8093 Zurich, Switzerland\\
  E-mail: \email{paolo.crivelli@cern.ch}}

\abstract{The understanding of the origin of dark matter has great importance for cosmology  and particle 
physics.
Several interesting extensions of the standard model dealing with solution of this problem motivate the concept of hidden sectors consisting of $SU(3)_C\times SU(2)_L\times U(1)_Y$ singlet fields. Among these models, the mirror  matter model is certainly one of the most interesting.
The model explains the origin of parity violation in weak interactions, it could also explain the baryon asymmetry 
of the Universe and provide a natural ground for the explanation of dark matter. The mirror matter could have  a portal to 
 our world through photon-mirror photon  mixing ($\epsilon$). This mixing would lead to orthopositronium  ($o-Ps$) to mirror orthopositronium oscillations, the experimental signature of which is  the apparently invisible decay of
  $o-Ps$.
In this paper, we describe  an experiment  to
 search for the decay  $o-Ps\to invisible $ in vacuum  by using
 a pulsed slow positron beam and a massive 4$\pi$ BGO crystal calorimeter.
The developed high efficiency positron tagging system, the low calorimeter
energy threshold and high hermiticity allow the expected 
sensitivity in mixing strength to be $\epsilon \simeq 10^{-9}$, which is more than one order of magnitude below the current 
Big Bang Nucleosynthesis limit and in a region of parameter space of great theoretical and phenomenological interest. The vacuum experiment with such 
sensitivity is particularly timely in light of the recent DAMA/LIBRA observations of the annual modulation signal consistent with a mirror type dark matter interpretation. 
}

\keywords{Positronium; Dark matter; Hidden sectors}

\begin{document}

\section{Introduction}

 Cosmological observations of  galactic rotational curves \cite{Borriello:2000rv} and the gravitational lensing \cite{Hoekstra:2002nf,Metcalf:2003sz} give strong evidence for the existence of dark matter  (see e.g. \cite{Bertone:2004pz} for an excellent review). In addition, the recent confirmation by the DAMA/LIBRA  experiment \cite{DAMALIBRA} of the annual modulation  signal observed  by the DAMA/NaI \cite{dama} seems  provide the first direct terrestrial experimental  observation of the existence of non-baryonic dark matter in our galactic halo. 
The need to explain these hints of the presence of dark matter 
provides one of the strongest indications for 
the existence of new physics beyond the Standard Model (SM) and 
the identification of the origin of dark matter is a task of 
enormous importance for both particle physics and cosmology.
At present, the most popular candidates for the 
(thermal-produced) dark matter are the so-called weakly interacting massive particles 
(WIMPs), which are e.g. lightest supersymmetric particles, Kaluza-Klein 
particles in universal extra dimension models, axions etc... 
However, despite of significant efforts the experiments looking for WIMPs 
lead so far to negative results, thus, pushing further possible WIMP searches  into  a very high-energy and/or high sensitivity frontiers, for a recent review see e.g. \cite{feng} and references therein.
%Supersymmetry is considered as one of the most attractive extension of the SM, mainly because it provides an elegant solution to the hierarchy problem \cite{Martin:1997ns} and enables  the couplings of the SM to evolve on a common scale in GUT theories \cite{Ellis:1991ri}. 
%The idea of a hidden sector is a generic feature of supersymmetric models at a weak scale.
%Supersymmetric-particles could be viable candidates for dark matter (even though that's not a compelling condition for their existence) if one forces R-parity to be conserved leading superpartners to annihilate or be created in pairs and, therefore, the lightest supersymmetric partner (LSP) to be stable.  
%Hoverer, the new results of DAMA/LIBRA gives us a hint that probably standard neutralinos models are not the solution of the the dark matter puzzle. In fact, such an explanation would contradict other higher threshold experiments like CDMS/Ge \cite{CDMS/Ge}, CDMS/Si \cite{CDMS/Si} and XENON10 \cite{XENON10}. of particle physics.  The new observations of dwarf spheroidal galaxies seem to indicate that non-cold dark matter (e.g., sterile neutrinos) needs to be considered seriously \cite{dm}.

An additional natural ground for the explanation of dark matter is provided by a class of interesting  theoretical models
 introducing the concept of ''hidden'' sectors consisting of $SU(3)_C\times SU(2)_L\times U(1)_Y$ singlet fields. 
It is worthwhile to note that, even in the SM, some of the matter fields are singlet  
under one or more of the colour and electroweak gauge groups. Thus, the extension to 
include a further sector which transforms under the new but not under the familiar gauge symmetries is not particularly exotic from a theoretical viewpoint. 
 
 The sensitivity of searches for the new singlet particles depends in detail on their  couplings and  mass scale,
 % However, it is clear that   direct sensitivity to a neutral 
%hidden sector will depend in detail on the couplings, which may be associated with higher 
%mass states charged under the SM that have been integrated out, as well as possible light 
%neutral states present at low energies.
% The energy scale at which new physics could  to be found, from general 
%arguments is usually associated to the high-energy frontier. 
e.g. if the mass scale of a hidden sector is too high,  
it is experimentally unobservable and indeed is hidden. Then the question arises: could the important
sensitive  searches for the hidden sectors be performed at low energy frontier ? 
The answer for this question is definitely positive. For example, there is a class of models with at least one additional $U(1)$ gauge factor where the corresponding hidden gauge boson could be light, or even massless \cite{mfok, holdom, mf}, for a recent review see  \cite{jr}.
 With respect to the mass and lifetime range for hidden sector 
states, the niche with masses in the range of a few MeV to a few GeV, with lifetimes of less than 1 second, is also quite intriguing as many of the most severe astrophysical and cosmological constraints that apply to lighter states are weakened or eliminated, while those from high-energy colliders are often inapplicable \cite{pospel}. 

Recent hints from terrestrial and  astrophysical anomalies, in particular first 
CDMSII result \cite{cdms} and annual modulations from DAMA/LIBRA at Gran Sasso \cite{DAMALIBRA, dama}, positron and/or electron-positron pair excess in cosmic ray from PAMELA \cite{pam} and ATIC \cite{atic} 
(however, the ATIC excess seems has not been confirmed by Fermi-LAT \cite{fermi})
 are found to be consistent with an interpretation in 
context of dark matter charged under new gauge fields \cite{dmreview}. Thus,
if those observations are really signals for DM they would indicate particularly exotic forms of DM. This suggests  that testing hidden sectors with the relatively light mass scale states in the MeV-to-GeV  range  weakly coupled to the SM model represents an intriguing possibility. 

Then it is important to ask  which hidden sector models  could be tested with the best experimental sensitivity? 
Certain classes of hidden sectors may be  tested, but it is clear that not  many generic scenarios, which are viable, could be strongly  constrained and even excluded on the base of precision 
measurements. 
Among numerous  candidates that have been discussed, one of the most promising, which could reconcile the  DAMA annual modulation signal with the results of the CDMS II experiment, is mirror type dark matter \cite{foot4,foot5}. 
In addition, the CoGeNT collaboration has recently reported a rising low energy spectrum in 
their ultra low noise germanium detector \cite{cogen}. This is particularly interesting as the 
energy range probed by CoGeNT overlaps with the energy region in which DAMA 
has observed their annual modulation signal. It has been recently shown \cite{foot6}, that the mirror dark matter 
candidate can simultaneously explain both the DAMA annual modulation signal 
and the rising low energy spectrum observed by CoGeNT. This constitutes a model 
dependent confirmation of the DAMA signal and adds weight to the mirror dark 
matter paradigm. 

In this paper, we show that the proposed search could result 
either in a strong evidence of the existence of dark matter of the mirror type, or completely 
exclude this type of new physics beyond the standard model.
It should be noted that mirror baryons are
naturally dark, stable and massive. Currently, it seems that this type of matter could also explain 
in a natural way the coincidence between visible and dark matter densities in the universe ($\Omega_B=0.044$ and $\Omega_{DM}=0.26$) \cite{berezhiani,Ciarcelluti:2008vs}. Interestingly, it could also provide an explanation of the controversial behaviour in galaxy cluster collisions \cite{silagadze}.
Furthermore, if mirror matter is present in our universe it would mean that Parity (spatial-inversion) is an unbroken symmetry of nature. 
This gives an exciting motivation for testing this model by a laboratory experiment at low energies. 

The rest of the paper is organized as follows. In Sec. \ref{sec:mirror}, we review the mirror model and work out briefly the mirror Higgs boson
physics at the CERN LHC, and in more details, the effect of oscillation of ordinary positronium to the mirror one and its experimental consequences.  We also present the calculations we performed in order to estimate the effect of matter and external fields on the oscillation probability. In Section \ref{sec:exptech}, we present the experimental technique. In Section \ref{sec:setup}, the design of the experiment and the detector components are described in detail. The simulations of the signal and background sources, as well as the expected sensitivity  are discussed in Section \ref{sec:Bkg_vac} and \ref{sec:sensitivity}, respectively. Section \ref{sec:conclusion} contains concluding remarks. 

\section{Mirror Matter model}\label{sec:mirror}

Mirror matter was originally discussed  by 
Lee and Yang \cite{ly} in 1956, after their discovery of parity violation  (for an excellent recent review on this subject see \cite{Okun:2006eb}). In order to save parity conservation they suggested that the transformation 
in the particle space corresponding to the space inversion  ${\bf x\to -x}$
should not be the usual transformation  {\it  P} but  {\it
  PR}, where  {\it  R} 
corresponds to the transformation of a particle (proton \cite{ly}) into a
reflected state in the mirror particle space.  
After the observation of parity non-conservation, Landau assumed \cite{dau}
that  {\bf \it  R=C}, i.e. he suggested to identify antiparticles with the 
mirror matter but then  {\bf \it CP} must be conserved, which we know is not 
the case.
The idea was further developed 
by A. Salam \cite{salam}, and was clearly formulated in 1966
as a concept of the mirror universe 
by Kobzarev, Okun and Pomeranchuk \cite{kop}. 
In their paper, they have shown that ordinary and mirror matter can 
communicate  predominantly through gravity and proposed that the 
mirror matter objects can be present in our universe.

Since that time, the concept of mirror matter has found many interesting 
applications and developments. In the 80's, it has been boosted by 
superstring theories with $E_8\times E_8^{'}$ symmetry, where the particles 
and the symmetry of interactions in each of the $E_8$ groups are identical.
Hence, the idea of mirror matter can be naturally combined in these models \cite{dubovsky}. 

Nowadays, mirror matter models exist in two basic versions. The symmetric 
version, proposed earlier, was further developed and put into a modern context 
by Foot, Lew and Volkas \cite{flv}. The asymmetric 
version was proposed by Berezhiani and Mohapatra \cite{bm}.

In the following we will concentrate on the symmetric model since it is the most interesting from a dark matter perspective and it could provide, as we will see, an experimental signature related to positronium. 
In the symmetric mirror model, the idea is 
that for each ordinary particle, such as the photon, electron, proton
and neutron, there is a corresponding mirror particle of 
exactly the same mass as the ordinary particle. 
R-parity interchanges the ordinary particles with the
mirror particles so that the properties of the mirror
particles completely mirror those of the ordinary particles.
For example, the mirror proton and mirror electron are stable and 
interact with the mirror photon in the same way in which the
ordinary proton and electron interact with the ordinary photons.
The mirror particles are unlikely to be produced
in laboratory experiments just because they couple very
weakly with the ordinary particles. In the modern language of gauge
theories, the mirror particles are all singlets under 
the standard $G \equiv SU(3)\otimes SU(2)_L \otimes U(1)_Y$
gauge interactions \cite{flv}. The mirror
particles interact with a set of mirror gauge particles,
so that the gauge symmetry of the theory is doubled,
i.e. the minimal gauge group of the new mirror model is 
$G_{SM} \otimes G'_{SM} \equiv SU(3)_C\otimes SU(2)_L \otimes U(1)_Y\otimes SU'(3)_C\otimes SU(2)'_L \otimes U'(1)_Y$ (the ordinary particles are, of 
course, singlets under the mirror gauge symmetry)~\cite{flv}.
 The gauge quantum numbers under $G_{SM} \otimes G'_{SM}$ for 
the usual and new fermion fields are 
\begin{eqnarray}
(L_L)^i\sim(1,2,-1)(1,1,0),~(L'_R)^i \sim (1,1,0)(1,2,-1) \nonumber \\
(e_R)^i \sim (1,1,-2)(1,1,0),~(e'_L)^i \sim (1,1,0)(1,1,-2) \nonumber \\
(Q_L)^i\sim (3,2, \frac{1}{3})(1,1,0),~(Q'_R)^i \sim (1,1,0)(3,2, \frac{1}{3})\nonumber \\
(u_R)^i \sim (3,1, \frac{4}{3})(1,1,0),~(u'_L)^i \sim (1,1,0)(3,1, \frac{4}{3})\nonumber \\
(d_R)^i \sim (3,1, -\frac{2}{3})(1,1,0),~(d'_L)^i \sim (1,1,0)(3,1, -\frac{2}{3})\nonumber \\
\end{eqnarray} 
with i the family index. 
In the left-right symmetric models parity is extended to a new type $Z_2$ 
discrete symmetry which transforms the left-handed field
to the right-handed one for the same fermion. However, the new 
 $Z_2$  parity symmetry that we can define now, is 
\begin{eqnarray}
{\bf x \leftrightarrow -x}, ~ t \leftrightarrow t,~G^\mu \leftrightarrow G'_\mu, ~W^\mu \leftrightarrow W'_\mu,    \nonumber \\
B^\mu \leftrightarrow B'_\mu,~L_L \leftrightarrow L'_R,    ~e_R \leftrightarrow e'_L, ~Q_L \leftrightarrow Q'_R,\nonumber \\
u_R \leftrightarrow u'_L, ~d_R \leftrightarrow d'_L
\label{zt}
\end{eqnarray} 
We see, that under $Z_2$ of (\ref{zt}),  the left-handed sector of the 
fermion field can transform to the right-handed sector of a 
different fermion field, namely, the {\em mirror fermion field}.
Thus, the $Z_2$ symmetry can be interpreted as 
a parity symmetry $(x\to -x)$, if the roles of left and right chiral fermion fields are 
interchanged in the mirror sector. 
Parity is conserved because the mirror particles experience
$V+A$ (i.e. right-handed) mirror weak interactions
while the ordinary particles experience the usual $V-A$ (i.e.
left-handed) weak interactions.  

An exact Lagrangian $Z_2$ symmetry interchanging ordinary and mirror particles 
is hypothesized, which means that all the couplings in the mirror sector are the 
same as in the ordinary sector. While, 
the ordinary and mirror particle sectors can interact between each other in a number of ways. The first is 
through gravitation, with immediate consequences for the dark matter problem and astro- 
physics \cite{blin}. 
Non-gravitational interactions can be induced through the mixing of colourless and 
neutral particles with their mirror counterparts. Neutrinos \cite{mn}, 
the photon \cite{glashow, gninenko, foot}, and 
the physical neutral Higgs boson \cite{ignasha, mirhiggs} can mix with the corresponding mirror states. Coloured 
and/or electrically charged particles are prevented from mixing with their mirror analogues 
by colour and electric charge conservation laws.

It is known \cite{flv} that there are two renormalizable and gauge invariant Lagrangian terms coupling the ordinary and mirror sector together: $U(1)_Y - U(1)'_Y$ 
gauge boson kinetic mixing, and ordinary-mirror Higgs scalar interactions, i.e. 
\begin{equation}
\frac{\epsilon}{2}F'_{\mu \nu}F^{\mu \nu}
\label{photon}
\end{equation}
and
\begin{equation}
\lambda \psi^\dagger_{SM}\psi_{SM}\psi^\dagger_{H}\psi_{H}
\label{hig}
\end{equation}

Based on this observation, one can extend the SM by 
doubling the ordinary fermion, gauge, and Higgs fields 
\cite{flv}. Thus, the new mirror fermions are natural singlets of 
the SM gauge group, and they (nucleus if there exists mirror  $SU(3)_C$) can be the candidates for dark matter. 

Mirror matter is invisible to us because it does not 
interact with ordinary photons and naturally constitutes a dark matter candidate. 
One should stress that the fact that our and mirror-sectors have the same micro-physics, 
does not imply that their cosmological evolutions should be the same. Indeed, if 
mirror particles had the same temperature in the early universe as ordinary ones, 
this would be conflict with Big Bang Nucleosynthesis (BBN). The BBN limit on the 
effective number of extra neutrinos implies that the temperature of the mirror sector 
$T'$ must be at least about twice smaller than the temperature $T$ of the ordinary sector 
allowing mirror baryons to be a viable candidate for dark matter. In particular, the 
mirror dark matter scenario would give the same pattern of Cosmic Microwave Background and Large Scale Structure
 as the standard CDM if $T'/T \lesssim  0.2$ \cite{bm,blin,ber1}. 
In addition, the baryon asymmetry of the Universe can be generated via out-of- 
equilibrium $B-L$ and $CP$ violating processes between ordinary and mirror particles 
\cite{ber2} whose mechanism could explain the intriguing puzzle of the correspondence between the visible and dark matter fractions in the Universe, naturally predicting the 
ratio $\Omega_{DM} /\Omega_B \simeq 5$ \cite{ber3}. 

\subsection{Higgs portal into mirror world at the CERN LHC}

In the mirror model every standard particle, 
including the physical neutral Higgs boson, is paired with a parity partner. The
interaction of Eq. (\ref{hig}) and  
unbroken  parity symmetry forces the mass eigenstate Higgs bosons to be 
maximal mixtures of the ordinary and mirror Higgs bosons:
\begin{equation}
H_\pm = \frac{H\pm H'}{\sqrt{2}}
\label{higgs}
\end{equation}
Each of these 
mass eigenstates will therefore decay 50\% of the time into invisible mirror 
particles. The total decay rate of $H_+$ or $H_-$ is the same as that for a SM physical neutral Higgs boson 
of the same mass. This may result in dramatic consequences for the LHC, making the significance of the 
Higgs signal at LHC lower due to decreasing of the {\em Signal/Background} ratio 
if the mass splitting is large compared  to the Higgs mass resolution at the LHC.
Note also that each mass eigenstate couples to ordinary particles with 
strength reduced by 1/2 compared to the coupling of the standard Higgs boson to those same particles \cite{ignasha}. 

Double Higgs peak observation for Eq. (\ref{higgs}), would give a clear and interesting signature for the Large Hadron 
Collider(LHC) which could thus establish the existence of the mirror world. 
However, for this effect to be observable the mass difference between the two 
eigenstates must be sufficiently large. The  cosmological 
constraints from Big Bang Nucleosynthesis on the mass difference parameter
have been studied in \cite{ignasha}. 
To summarize, one can see that the Higgs sector may play an important role in detecting mirror particles, 
which can be the candidates 
of dark matter and appear as missing energy in the detectors at the LHC 
\cite{ignasha, mirhiggs}. Another intriguing possibility discussed in the next Sec. is related to the possible observation
of orthopositronium to mirror orthopositronium oscillations.

\subsection{Positronium portal into mirror world}

\begin{figure}[h!]
\begin{center}
\includegraphics[width=0.3\textwidth]{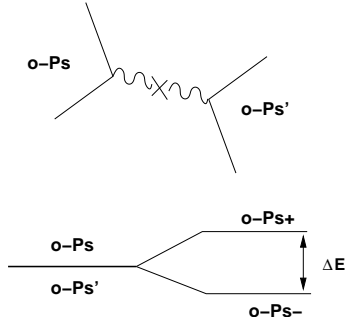}
 \caption{\em The double degeneracy between orthopositronium mass eigenstates
of ordinary (\ops ) and mirror (\mops ) is broken when a small mixing term is included. }
\label{mixing}
\end{center}
\end{figure}

Positronium ($Ps$), the positron-electron bound state,
 is the lightest known atom, which is bounded and self-annihilates 
through the same, electromagnetic interaction. At the 
current level of experimental and theoretical precision this is  
the only interaction present in this system  \cite{karshenboim2004}. 
This feature has made positronium an ideal system for  
testing the accuracy of QED calculations 
for bound states, in particular for the triplet ($1^3S_1$)
state of $Ps$, orthopositronium ($o-Ps$). 
Due to the odd-parity under
C-transformation,  $o-Ps$ decays
predominantly into three photons. 
Due to the phase-space and  additional
$\alpha$ suppression factors, as compared with the singlet ($1^1S_0$) state (para-positronium), the "slowness" of $o-Ps$ decay rate  gives an enhancement factor $\simeq 10^3$ in sensitivity to an admixture of 
new interactions which are not accommodated in the Standard Model  \cite{wsandre, psrev}.
Glashow  realized that
the orthopositronium system provides one sensitive
way to search for the mirror matter \cite{glashow}. 
 Glashow's idea is that if a small kinetic mixing between ordinary and mirror
photons exists \cite{holdom}, it would mix ordinary and mirror 
orthopositronium, leading to maximal orthopositronium -
mirror orthopositronium oscillations (see Fig.~\ref{mixing}). Since mirror \mops\
decays predominantly into three mirror photons these oscillations would
result in \invdecay\ decays in vacuum.
Photon-mirror photon kinetic mixing
is described by the interaction Lagrangian density

\begin{equation}
L = \epsilon F^{\mu \nu} F'_{\mu \nu},
\label{ek}
\end{equation}
where $F^{\mu \nu}$ ($F'_{\mu \nu}$) is the field strength 
tensor for electromagnetism (mirror electromagnetism).

Together with the Higgs- mirror Higgs quartic couple $\lambda \phi \phi^{\sp\dagger} \phi' \phi'^{\sp\dagger} $,
 these are the only renormalizable and gauge invariant terms that can be added to the SM Lagrangian. 
The effect of ordinary photon - mirror photon kinetic mixing
is to give the mirror charged particles a small electric
charge \cite{flv,holdom,glashow}. That is, they couple to ordinary photons with
charge $2\epsilon e$\footnote{Note, that the direct experimental
bound on $\epsilon$ from searches for `milli-charged' particles
is $\epsilon \stackrel{<}{\sim} 10^{-5}$
\cite{prinz}.}.

Orthopositronium is connected via a one-photon
annihilation diagram to its mirror version (\mops )~\cite{glashow}.
This breaks the degeneracy between \ops\  and \mops\ so that
the vacuum energy eigenstates are $(o-Ps + o-Ps' )/\sqrt{2}$ and 
$(o-Ps - o-Ps' )/\sqrt{2}$,
which are split in energy by
\begin{equation}\label{eq:enerysplitting}
\Delta E = 2h\epsilon f, 
\end{equation}

where $f = 8.7\times 10^4$ MHz is the contribution to the
ortho-para splitting from the one-photon annihilation diagram
involving \ops\ \cite{glashow}. Assuming a mixing strength of $4\times 10^{-9}$ (as suggested by the DAMA results), one obtains an energy splitting of $\Delta E = 2.9\times 10^{-12}$ eV.
Thus, the interaction eigenstates are maximal combinations
of mass eigenstates which implies that \ops\ oscillates
into \mops\ with a probability: 
\begin{equation}
P(p-Ps \to o-Ps' ) = \sin^2 \Omega t, 
\end{equation}
where $\Omega = 2\pi\epsilon f$.

The simplest case of $o-Ps \to o-Ps' $ 
oscillations in vacuum~\cite{glashow} leads to an {\it apparent} increase in the decay
rate because the mirror decays are not detected. The number of \ops\, $N$ satisfies
\begin{equation}
N = \cos^2 \Omega t \cdot e^{-\Gamma_{SM} t}
\simeq exp [-t(\Gamma_{SM} + \Omega^2t)],
\end{equation}
where
$\Gamma_{SM}$ is the Standard Model 
decay rate of $o-Ps$ \cite{ADKINS-4,Tokyo_2003,Michigan_2003}. 
Thus $\Gamma^{eff} \approx \Gamma_{SM}(1 + \Omega^2/\Gamma_{SM}^2)$ leads to a branching ratio of:

\begin{equation}\label{eq:br_mirror_vacuum}
Br(o-Ps\to invisible) = \frac{2(2\pi \epsilon f)^2}{\Gamma^2_{SM} + 
4(2\pi \epsilon f)^2}.
\end{equation}

The above calculation is not applicable to
an experiment performed with a cavity confining the positronium,
 because in this case
the collision rate is not zero and the 
loss of coherence due to the collisions must be included
in the calculation~\cite{gninenko,foot}. In those papers, the collision rate was assumed to be much larger than the decay rate $\Gamma_{coll} \gg \Gamma_{SM}$ \cite{foot}, thus, an approximate solution for this case was found. In the proposed experiment, we are interested to see the effect on the oscillation probability for 1-2 collisions per lifetime. Furthermore, external electric or magnetic field induce additional splitting of \ops\ and \mops\ states affecting the probability $P(o-Ps \to o-Ps' )$.

Only the second order Stark shift contributes to the positronium in the ground state. The energy shift can be calculated in the same way as hydrogen and is given by \cite{bethesalpeter}:
\begin{equation}\label{eq:starkshift}
\Delta_{stark} = -\frac{1}{2} \alpha_0 \epsilon_0 E^2,
\end{equation}
where $4.54 \alpha_0=\pi a_0^3$ is the polarizability of \ops\  ($a_0=0.1$ nm is the Bohr radius of positronium), $\epsilon_0$ is the vacuum permittivity and $E$ is the electric field. 
Magnetic fields affects only the triplet state with quantum number m=0. For the m=$\pm$1 states, the Zeeman effect is zero because the magnetic moments for positron and electron are opposite. The energy contribution for m=0 can be calculated with:   
\begin{equation}\label{eq:zeemanshift}
\Delta_{zeeman} = -\frac{A{\hbar}^2}{2}+\sqrt{(\frac{A{\hbar}^2}{2})^2+{\hbar}^2g_1^2B_0^2}
\end{equation}
where $A=4.92\times 10^{25}$ eV$^{-1}$s$^{-2}$ is the hyperfine splitting constant of the ground state, $g_1=2.81\times 10^{10}$ T$^{-1}$s$^{-1}$ the gyromagnetic ratio, $B_0$ the magnetic field. 

\begin{figure}[h!]
\hspace{.0cm}\includegraphics[width=.5\textwidth]{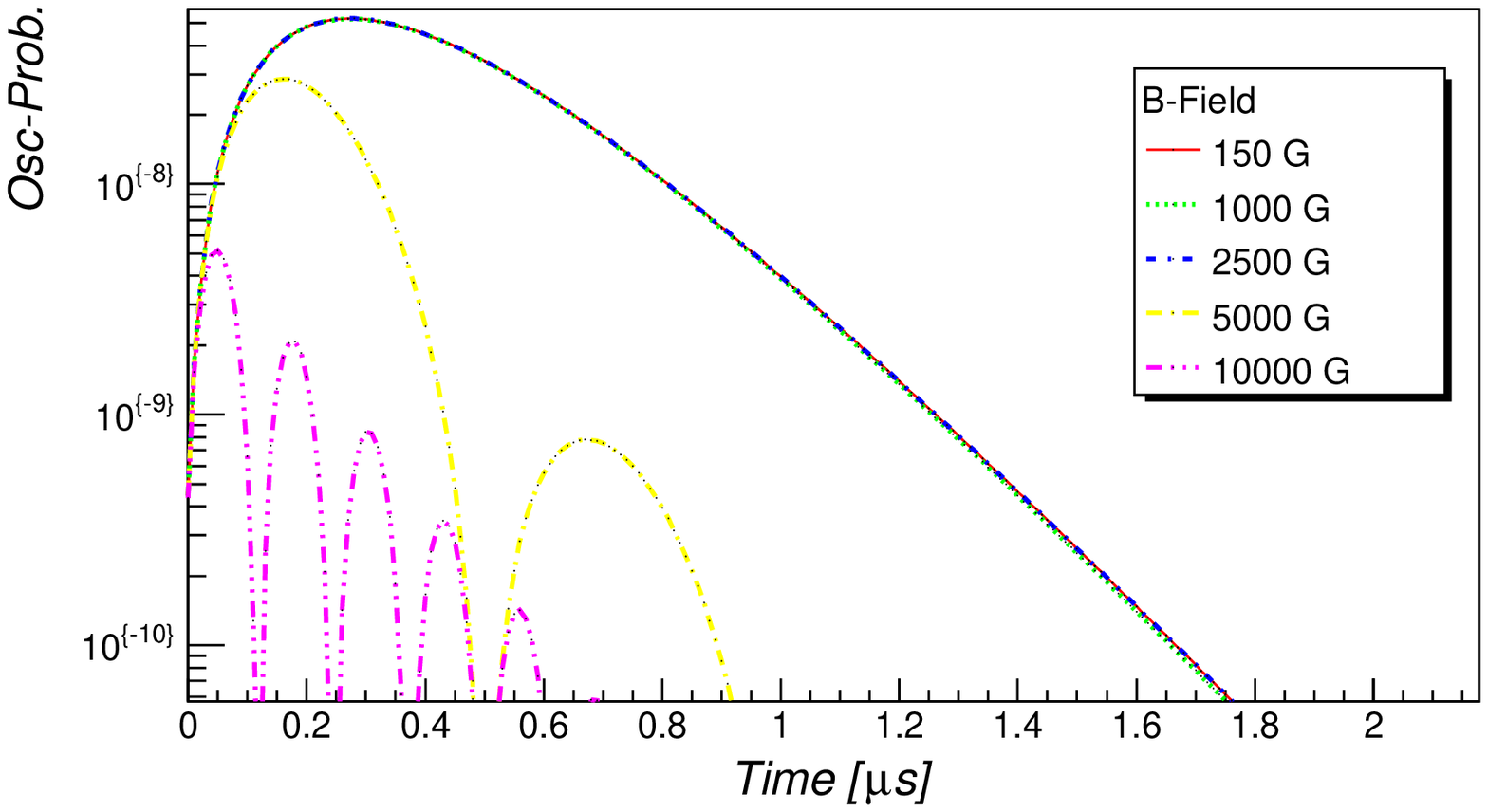}
\hspace{.0cm}\includegraphics[width=.5\textwidth]{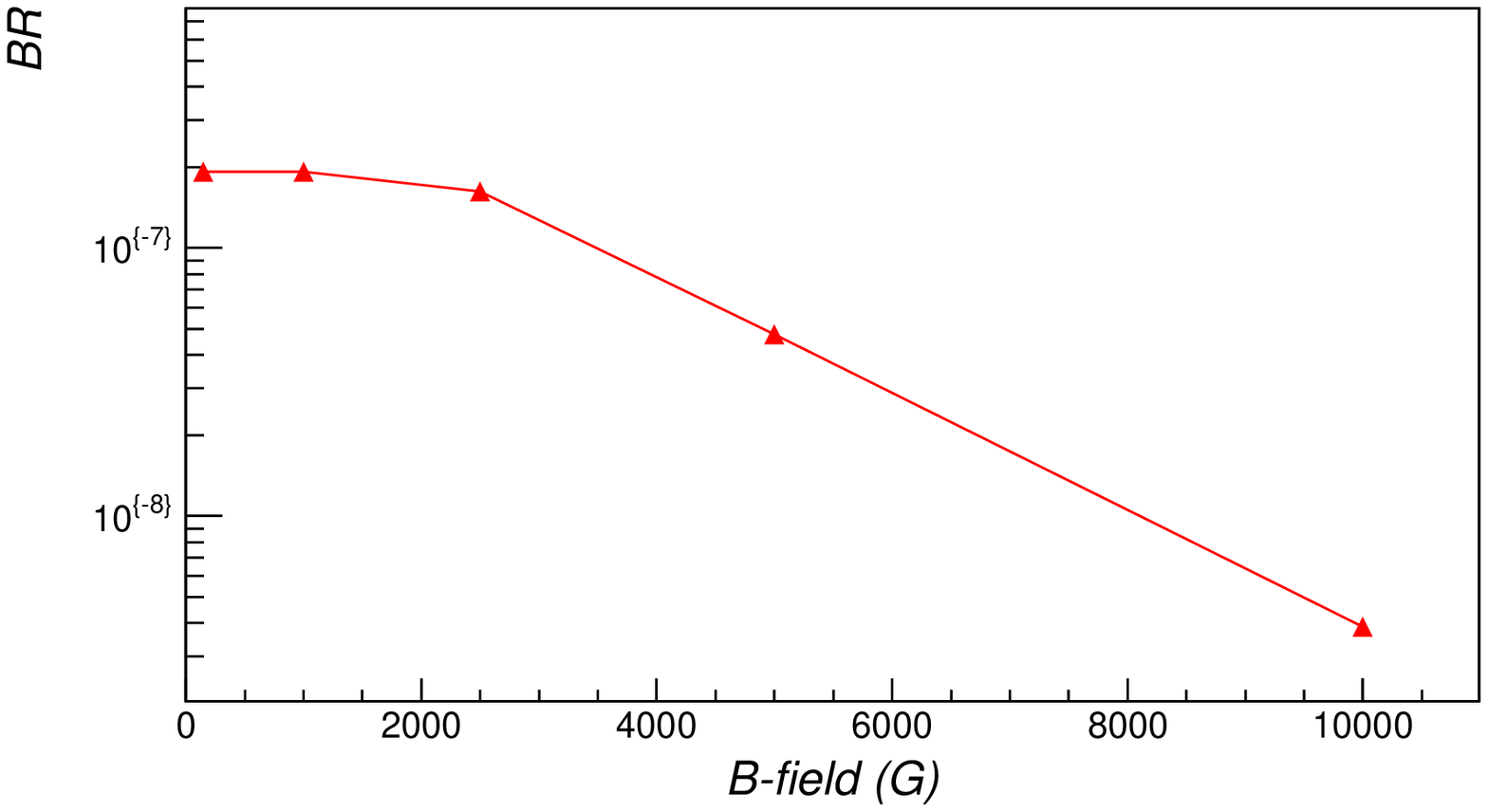}
 \caption{\em Oscillation probability and branching ratio for \binvdecay\ as a function of the magnetic field. The electric field was set to 5 kV/cm. }
\label{BRvsBField}
\end{figure}

\begin{figure}[h!]
\hspace{.0cm}\includegraphics[width=.5\textwidth]{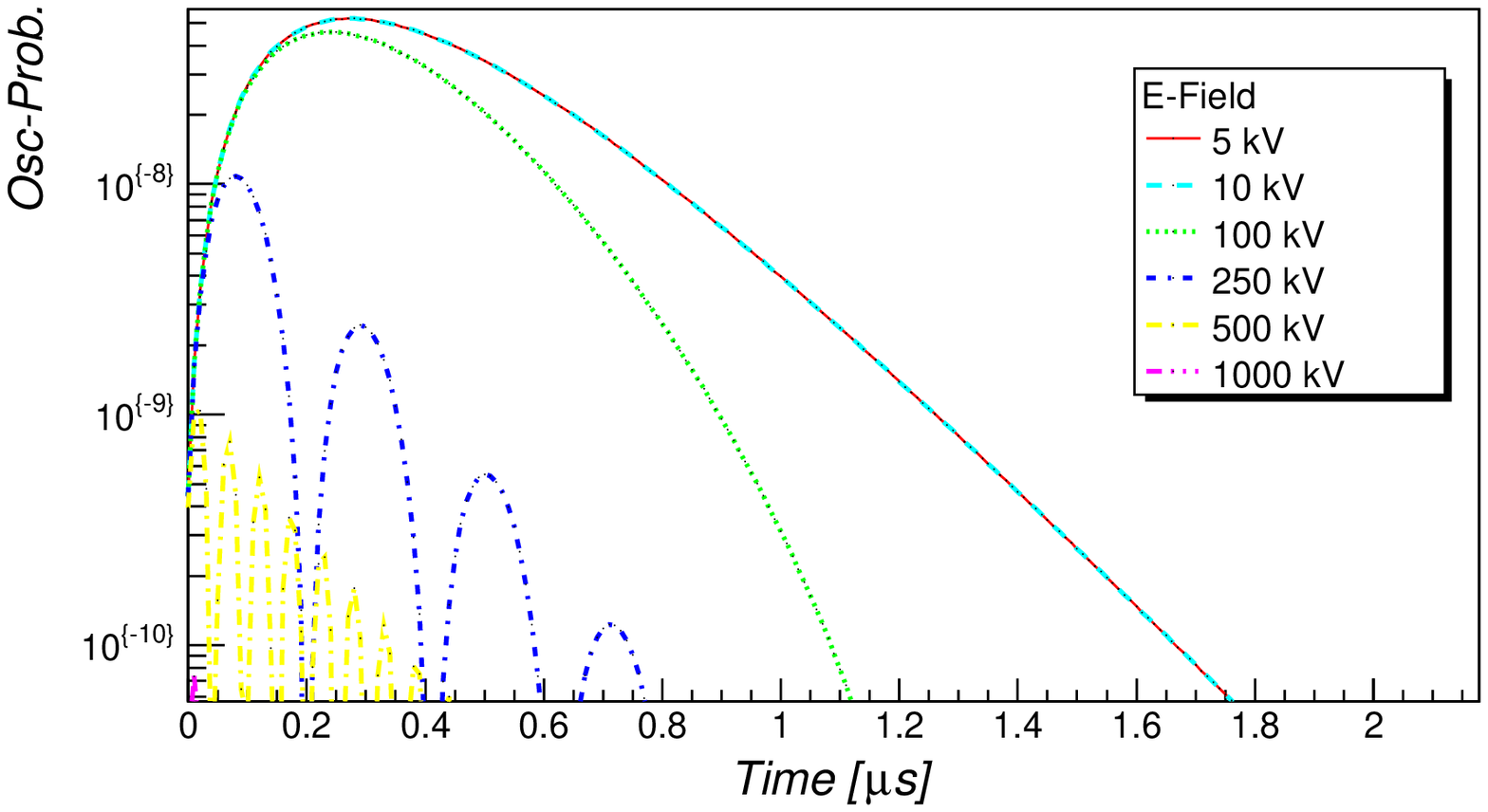}
\hspace{.0cm}\includegraphics[width=.5\textwidth]{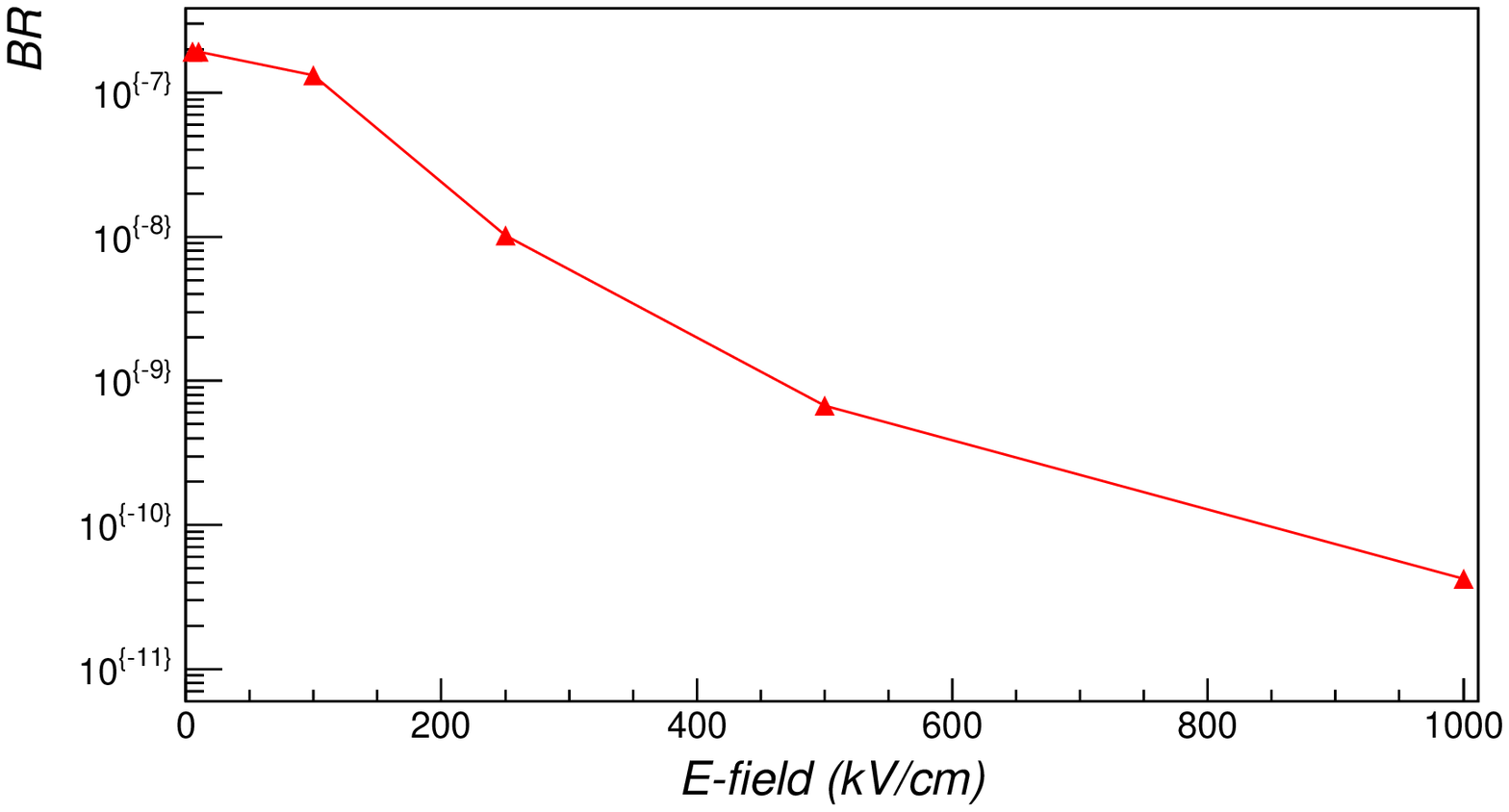}
\caption{\em Oscillation probability and branching ratio for \binvdecay\ as a function of the electric field. The magnetic field was set to 150 G. }
\label{BRvsEField}
\end{figure}

\begin{figure}[h!]
\begin{center}
\hspace{.0cm}\includegraphics[width=.5\textwidth]{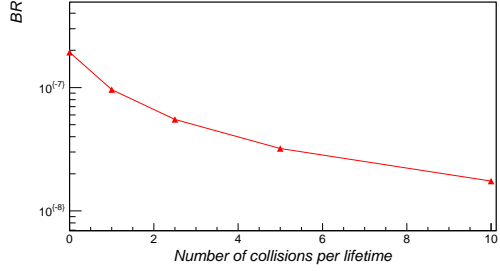}
\end{center}
 \caption{\em Branching ratio for \binvdecay as a function of the number of collisions per lifetime of \ops\ with the cavity walls. The electric field was set to 5 kV/cm and the magnetic field was 150 G.}
\label{BRvsNcoll}
\end{figure}

To include these effects in the calculation, we use the density matrix approach.
The Heisenberg's equation of motion
\begin{equation}\label{eq:Heisenberg}
\dot{\rho} = \frac{1}{i\hbar}[H,\rho]+\frac{d}{dt}\rho_{rel}
\end{equation}
can be solved to find the density matrix 
\begin{equation}
\rho(t)=\left( \begin{array}{cc}
\rho_{11}(t) & \rho_{12}(t)\\
\rho_{21}(t) & \rho_{22}(t)   
\end{array} \right) 
\end{equation} 
that describes the evolution of the states with time.
The hamiltonian H of the system is given by:
\begin{equation}\label{eq:HoPsMoPs}
H= \left( \begin{array}{cc}
\frac{1}{2}(\Delta+\omega_{12})\hbar & -\Omega \hbar\\
-\Omega \hbar & -\frac{\omega_{12} \hbar}{2}  \end{array} \right)
\end{equation}
where $\Delta=\Delta_{stark}+\Delta_{zeeman}$ is the shift of the ground state introduced by magnetic and electric fields. The effect of those fields on \mops\ can be neglected because of the very weak coupling. The term $\omega_{12}$ is the splitting between the two vacuum eigenstates.
To describe all the processes (decays and collisions) that return the ensemble in thermal equilibrium,
i.e. destroy the coherence of the oscillation, we use the relaxation term $\rho_{rel}$. Its derivative is equal to:
\begin{equation}\label{eq:rhorel}
\frac{d}{dt}\rho_{rel}= \left( \begin{array}{cc}
-\gamma_1 \rho_{11}(t) & -(\frac{\gamma_1+\gamma_2}{2}+\gamma_{coll})\rho_{12}(t)\\
 -(\frac{\gamma_1+\gamma_2}{2}+\gamma_{coll})\rho_{21}(t) & -\gamma_2 \rho_{22}(t)   \end{array} \right)
\end{equation}
where the constants $\gamma_1$ and $\gamma_2$ are the decay rate in vacuum of oPs and oPs', respectively. 
 The term $\gamma_{coll}$ is the collision rate of \ops\ with the cavity walls.
Substituting (\ref{eq:HoPsMoPs}) and (\ref{eq:rhorel}) in Eq. (\ref{eq:Heisenberg}), one obtains the four differential equations:
\begin{eqnarray}\label{eq:diffeqs}
 \dot{\rho}_{11}(t) &=& -\gamma_1 \rho_{11}(t)-\frac{i}{\hbar}(\Omega\hbar \rho_{12}(t) - \Omega\hbar \rho_{21}(t)) \nonumber \\
\dot{\rho}_{22}(t) &=& -\gamma_2 \rho_{22}(t)-\frac{i}{\hbar}(-\Omega\hbar \rho_{12}(t) + \Omega\hbar \rho_{21}(t)) \nonumber\\
\dot{\rho}_{12}(t) &=& -(\frac{\gamma_1+\gamma_2}{2}+\gamma_{coll}) \rho_{12}(t) \nonumber \\
 & & -\frac{i}{\hbar}\big(\Omega\hbar \rho_{11}(t)+\frac{1}{2}\omega_{12}\hbar \rho_{12}(t) + \frac{1}{2}(\Delta+\omega_{12})\hbar \rho_{12}(t) - \Omega\hbar \rho_{22}(t)\big) \nonumber \\ 
\dot{\rho}_{21}(t) &=& -(\frac{\gamma_1+\gamma_2}{2}+\gamma_{coll}) \rho_{21}(t) \nonumber \\ 
& & -\frac{i}{\hbar}\big(-\Omega\hbar \rho_{11}(t)-\frac{1}{2}\omega_{12}\hbar \rho_{21}(t) - \frac{1}{2}(\Delta+\omega_{12})\hbar \rho_{21}(t) + \Omega\hbar \rho_{22}(t)\big)
\end{eqnarray}
that can be solved numerically  (we used Mathematica \cite{math} for this purpose). At t=0 we start with a pure \ops\ state, thus, we set the  initial conditions: $\rho_{11}(0)=1, \rho_{12}(0)=0, \rho_{21}(0)=0, \rho_{22}(0)=0$.
 We assumed that without the coupling the vacuum eigenstates of \ops\ and \mops\ are degenerate and thus we set $\omega_{12}=0$ in the calculation. To find the numerical solution, we also used $\gamma_1=\gamma_2=1/142$ ns$^{-1}$ implying that \ops\ and \mops\ have the same lifetime in vacuum.
 To find the branching ratios (BR) one has to integrate over t $\rho_{11}$ and $\rho_{22}$, thus the BR is:
\begin{equation}\label{eq:BRInt}
BR=\frac{\int_0^t\! \rho_{22}(x) \, dx}{\int_0^t\! \rho_{11}(x) \, dx+\int_0^t\! \rho_{22}(x) \, dx}
\end{equation}
where the upper limit of the integration can be chosen as the data acquisition gate for the calorimeter. Assuming $\epsilon=4\times10^{-9}$ as suggested by the DAMA/LIBRA results, one can calculate the oscillation probability ($\rho_{22}(t)$)  and the branching ratios for \ops\ into \mops\  for different values of the EB-fields. The results are shown in Figs. \ref{BRvsBField}-\ref{BRvsEField}. As one can see, the branching ratio is not affected by the magnetic field (100 G) and by the electric field (less than 10 kV/cm) we are planning to use in the experiment. The effect of the collision is shown in \ref{BRvsNcoll}. As expected, solving Eq. (\ref{eq:BRInt}) for zero fields and no collisions leads to the same value of the BR obtained with Eq. (\ref{eq:br_mirror_vacuum}).

\subsection{Experimental evidence for Dark Matter of Mirror type}\label{sec:mirror}
 At present, there is some experimental evidence
that mirror matter could exist, coming from cosmology as well as from
the neutrino physics \cite{foot3}. Foot discussed implications of 
the DAMA experiment for mirror matter-type dark matter, 
which is coupled to ordinary matter through the interaction of Eq. (\ref{ek})
 \cite{foot1,foot4}.
 It has been shown that the annual modulation signal measured  by the DAMA/NaI experiment \cite{dama} can be
       explained by mirror matter-type dark matter if the photon-mirror photon
mixing strength is in the region 
\begin{equation} 
\epsilon  \simeq 4 \times 10^{-9}.
\label{ft1}
\end{equation}

\begin{figure}[htb!]
\begin{center}
\hspace{.0cm}\includegraphics[width=.6\textwidth]{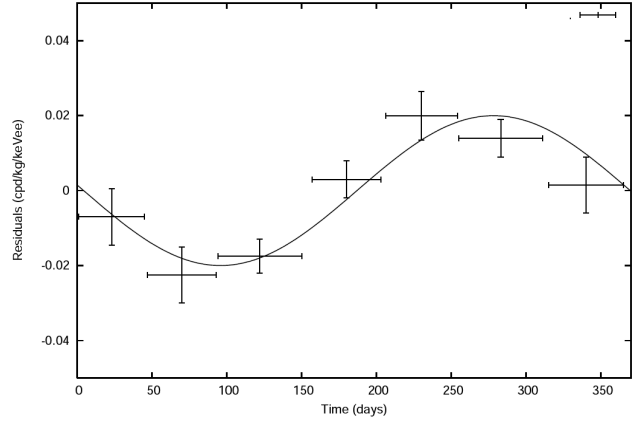}
\end{center}
\caption{\em The points are the annual modulation observed by DAMA/NaI. The line is the prediction of this modulation for mirror matter \cite{foot4}.}
\label{tof}
\end{figure}

Those results have been recently confirmed by the DAMA/LIBRA experiment with a 8.2$\sigma$ significance \cite{DAMALIBRA}. It was also pointed out \cite{foot5} that the very recently published results of the CDMSII/Ge final exposure \cite{cdms} with two events in the signal region with recoil energies close to the threshold can be explained by the interactions of mirror nuclei in the detector. Mirror matter is a promising candidate for dark matter since it can reconcile the null results from the other higher threshold experiments like CDMS/Si \cite{CDMS/Si} and XENON10 \cite{XENON10}.     
 Interestingly, this value of $\epsilon$ is also 
consistent with all other known experimental and cosmological bounds, 
including SN1987a\footnote{The SN1987a limit $\epsilon < 10^{-9.5}$ 
obtained in Ref.~\cite{kmt} is actually much weaker. 
For a more detailed discussion of this and other constraints see
  Ref.~\cite{fiv}}
 and the standard Big Bang Nucleosynthesis (BBN) 
bound  \cite{cg}. It was also confirmed that photon-mirror photon mixing of this magnitude is consistent as well with the more stringent constraints from cosmic microwave background measurements and large scale structure considerations \cite{ciarceluttiPLB2009}.
 It is also in the  range of naturally small 
$\epsilon$-values motivated by grand unification models \cite{berezhiani}.

If $\epsilon$ is
as large as in Eq. (\ref{ft1}),
the branching ratio $ Br(o-Ps\to invisible)$ for the invisible decay of 
orthopositronium in vacuum can be found with Eq. (\ref{eq:br_mirror_vacuum}) and is of the order: 
\begin{equation}
Br(o-Ps\to invisible)\simeq 2\times 10^{-7}.
\label{ft2}
\end{equation}
For comparison,  the BBN limits \cite{cg} deduced from the successful 
prediction of the primordial $^4$He abundance are
\begin{equation}\label{bbn1}
\epsilon < 3\times 10^{-8}
\end{equation}
and 
\begin{equation}
Br(o-Ps\to invisible) < 10^{-5}
\label{bbn2}
\end{equation}
respectively.

Given the indications for the mirror world,
coming from dark matter~\cite{foot1} and the neutrino physics 
anomalies~\cite{foot3,venj}, as well as the intuitive expectation
that nature could be left-right symmetric, 
it is obviously important to  determine experimentally
whether orthopositronium is a window to the mirror world or not. 

\section{Experimental technique}\label{sec:exptech}  

The experimental signature of \invdecay\ decay is the apparent disappearance of the energy  $2m_e$ expected in ordinary decays in a hermetic calorimeter surrounding the \ops\ formation target. Therefore, the occurrence of the $o-Ps\to o-Ps' \to invisible$ conversion would appear as an excess of events with zero-energy deposition in the calorimeter above those expected either from Monte Carlo prediction of the background or from direct background measurements.

The experiment presented here is based on the slow positron beam used to form \ops\ in a vacuum cavity combined with the BGO calorimeter used in our previous search for \invdecay\ decays \cite{oPsInv}, see also \cite{opsmos,opsjap}. A preliminary design of the apparatus has been presented in Ref. \cite{Gninenko:2004ft}. In the present work, more detailed 
simulations of the experimental setup, an improved positron tagging scheme, the selected positronium formation target and a better understanding of 
all possible background sources and sensitivity of the proposed search are presented.

The great advantages of this approach compared to the one, where we produced \ops\ in an aerogel target \cite{oPsInv}, are listed here:
\begin{itemize}
\item Compared to our previous experiment, a factor $10^4$ more statistics can be collected with the same number of positrons. In the thin SiO$_2$ films that we plan to use as a target, 10 times more \ops\ is produced per implanted positron (see Section \ref{subsec:target}) compared to the aerogel. Furthermore, there is no need to apply cuts for the 1.27 MeV photon selection and for the fiber energy deposition that reduced the number of events to less than 1\% with respect to the number of positrons emitted from the source \cite{oPsInv,thesis}. 
%Hence, the same number of \ops\ decays can be collected in one day instead of four months.
\item In the previous experiment, the main contribution to the 12\% inefficiency for the detection of events that gave a trigger, arose from the overlap of close in time annihilation events (so called pileup events). In the beam based experiment described here, after every 300 ns bunch the chopper shuts off the positron flux for 3 $\mu$s. Therefore, the efficiency for signal detection will be close to 100\% (see Section \ref{sec:sensitivity}).
%In addition, the threshold defining a zero-energy event in the ECAL will be smaller than the 80 keV used previously.    
\item  The suppression of the $\ops \to \ops^{'}\to invisible$ conversion due to the decoherence caused by the interaction of \ops\ with matter will be minimized. The number of collisions with the walls of the vacuum cavity undergone by the \ops\ during its lifetime is at least a factor $10^4$ smaller than in the aerogel pores. Since the branching ratio for the $\ops \to \ops^{'} \to invisible$ decay is approximately inversely proportional to the number of the \ops\ collisions (see Fig. \ref{BRvsNcoll}), the sensitivity on the mixing strength will be a factor 100 better for the same statistics.
\item In case an excess of events above the MC expectation for the background is observed, this experiment offers a unique and essential feature: one can cross check experimentally if it comes from signals. More precisely, by changing the number of collisions one can modify the oscillation probability while the background remains the same. We thought about two different possibilities: 
\begin{enumerate}
\item taking two runs at different positron implantation energies. From 3 to 5 keV the mean velocity of the created \ops\ increases by about a factor of two, thus, the collision rate with the walls is 2 times bigger and the signal is suppressed by the same factor (see Section \ref{subsec:target}). 
\item the same result can be achieved varying the length of the cavity confining positronium (see Section \ref{subsec:cavity}) while keeping the implantation energy of the positron fixed. 
%\item in a similar way small variations of gas pressure result in larger peak variations at zero energy due to the damping of
 %$\ops\to $ oscillations. However, this method could introduce a source of background associated with electrons and ions due to ionization of the residual gas atoms by positrons that needs to be further investigated. 
\end{enumerate}
 
\end{itemize}

However, compared to the previous experiment, there is a clear disadvantage: the calorimeter must be mounted outside the vacuum chamber so that the vacuum pipe introduces a loss of the photon energy. Nevertheless,  simulations show that with an aluminum pipe of 1 mm thickness the sensitivity of the experiment will be at a level of 10$^{-7}$ (see Sections \ref{subsec:cavity} and \ref{sec:sensitivity}).

\section{The setup}\label{sec:setup}

\begin{figure}[h!]
\includegraphics[width=1.0\textwidth]{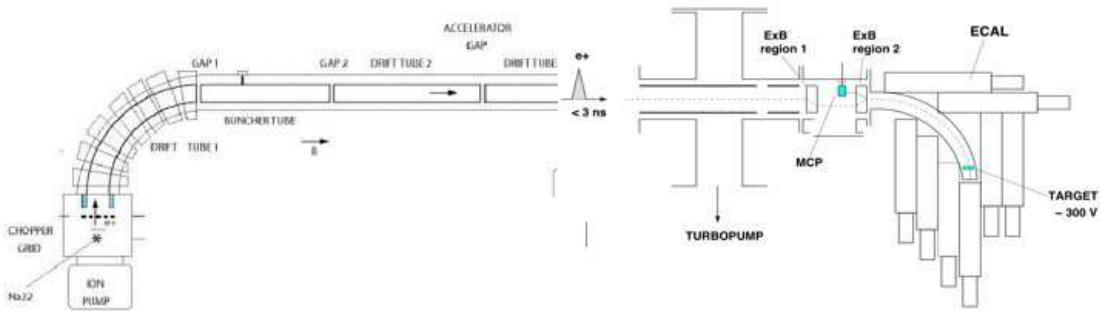}
\caption{\em Schematic view of the setup. The main parts of the system are: The bunched slow positron beam, the secondary electron tagging system, the target chamber and the $\gamma$ detector.}
\label{BeamInvSchem}
\end{figure}

The experiment is designed with the goal to observe the  $\ops \to invisible$ decay if its branching ratio is of the order of $10^{-7}$ (see Section \ref{sec:sensitivity}). 
According to this requirement, the apparatus has several distinct and separated parts (Fig. \ref{BeamInvSchem}) that will be described in detail in the following sections:
\begin{enumerate}
\item the bunched slow positron beam (Section \ref{subsec:beam})
\item the target for efficient \ops\ production near thermal energy (Section \ref{subsec:target}) 
\item the vacuum cavity to confine the \ops\ (Section \ref{subsec:cavity}) 
\item the positron appearance tagging system with a high S/N ratio, based on a high performance micro-channel-plate (MCP) described combined with the positron bunching (Section \ref{subsec:tagging})
\item the gamma detector, an almost 4$\pi$  BGO crystal calorimeter (ECAL) surrounding the vacuum cavity for efficient detection of annihilation photons to search for invisible \ops\ decays (Section \ref{subsec:InvDet}). 
\end{enumerate}

\subsection{ The slow positron beam}\label{subsec:beam}
\begin{figure}[h!]
\begin{center}
\includegraphics[width=.6\textwidth]{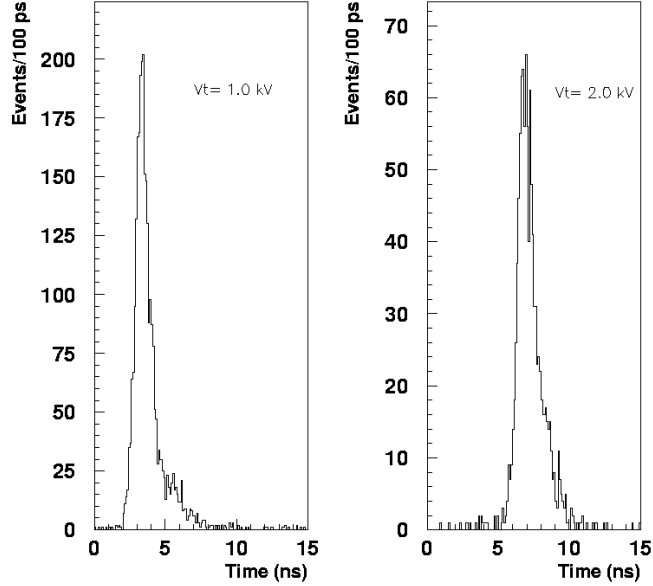}
\caption{\em Timing between the secondary electrons detected in the MCP and the bunching pulse of the beam for 1 and 2 keV implantation energy of the positrons.}
\label{MCPBunchingCoincidence}
\end{center}
\end{figure}

The details and performance of the ETHZ positron beam are presented in \cite{beamNIMA}. The beam is designed to operate in two modes, i.e. there are two different ways of tagging the positrons (or positronium):
\begin{enumerate}
\item Detection with a MCP of the secondary electrons emitted when the positrons hit the target. 
\item Positron bunching: an initial pulse of 300 ns is compressed to the target region into a 2 ns wide pulse. 
\end{enumerate}
%As presented in Chapter \ref{Ch:Beam},
In the first mode of operation, the secondary electrons (SE) produced by the positrons (about 25000 e$^+$/s) hitting the target are accelerated to 1-10 keV by the same voltage (applied to the target relative to the grounded transport tube) that is used to implant the positrons in the positronium converter (target). The secondary electrons are then transported in the backward direction by the same magnetic field, which guides the positrons. 
%as shown in Figure \ref{traj}. The electrons move along the magnetic field line in spirals and are deflected to the micro-channel plate (MCP) region by the $E\times B$ filter.
The two operation modes described above can be combined. This is done by studying the delay between the detection of the SE at the MCP with respect to the arrival time of the positron at the target: the trigger for the positron tagging can be formed by a coincidence of the pulse from the MCP and the signal from the pulsed beam (see Fig. \ref{MCPBunchingCoincidence}). This is a key point for the experiment because of the requirement to have the highest possible signal-to-noise (S/N) ratio in order to suppress fake triggers. With a fake trigger we mean that the trigger is not correlated with the presence of a positron in the \ops\ formation cavity. In this case, no annihilation photons would be detected and,  as a consequence, this event would be mis-identified as an invisible decay. The requirement of this coincidence suppresses:
\begin{itemize}
\item  the background generated from electrons and ions produced by ionization of the residual gas atoms by positrons. 
\item the accidentals due to the MCP noise to a level of 10$^{-5}$.
\end{itemize}

This fundamental step of the experiment has been already tested successfully and the result is shown in Figure \ref{MCPBunchingCoincidence}.
As will be explained in Section \ref{subsec:tagging}, to reach the required confidence level for the presence of a positron in the formation cavity, it is necessary to add an additional condition to the trigger scheme.

\subsection{ $\ops$ production target}\label{subsec:target}
\begin{figure}[h!]
\begin{center}
\hspace{.0cm}\includegraphics[width=.6\textwidth]{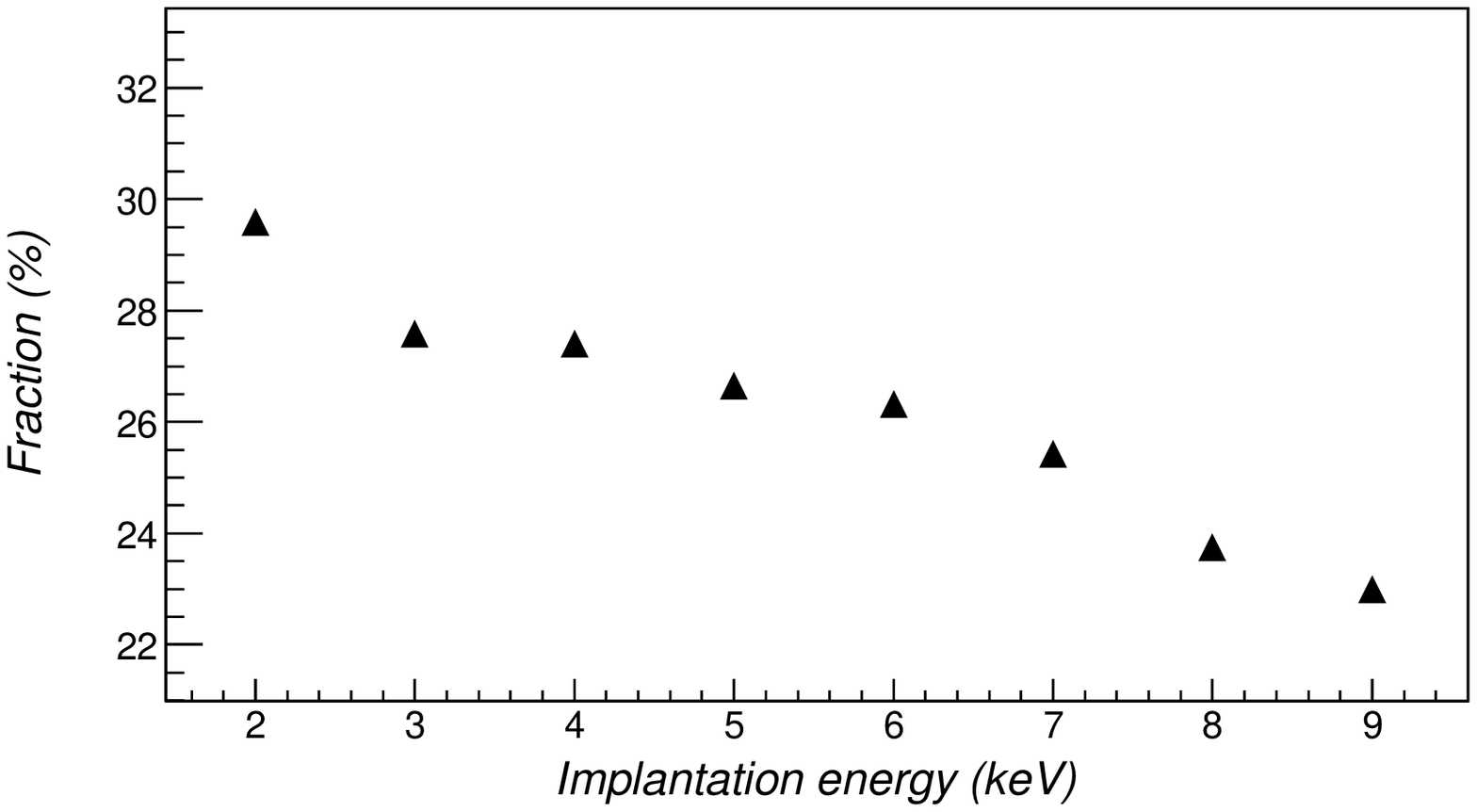}
\hspace{.0cm}\includegraphics[width=.6\textwidth]{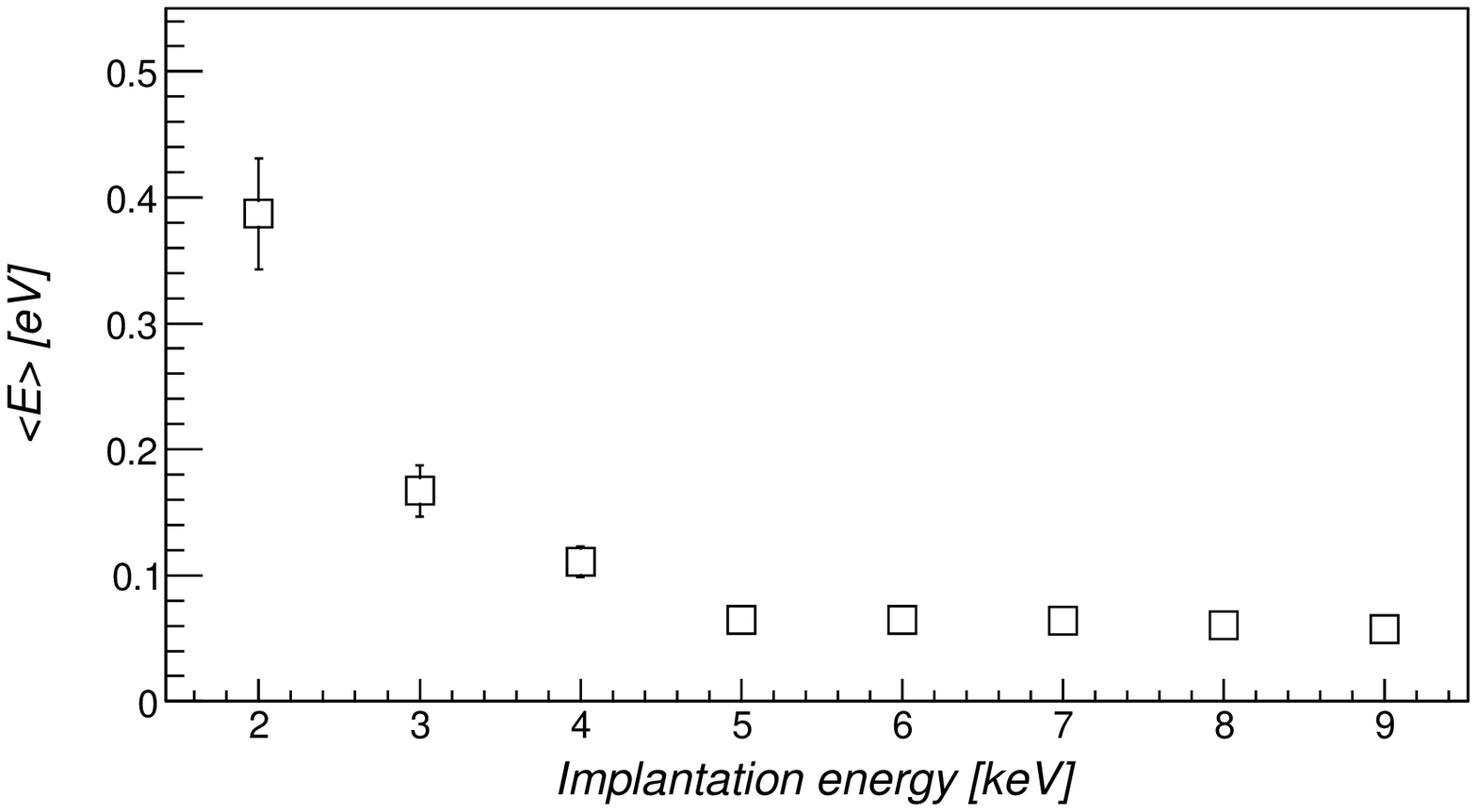}
%\hspace{.0cm}\includegraphics[width=.8\textwidth]{142_F127.eps}
\caption{\em (Top plot) Total fraction of \ops\ produced in the SiO$_2$ targets as a function of the positron implantation energy measured with positron annihilation lifetime spectroscopy. (Bottom plot) Emission energy of \ops\ in vacuum measured with time-of-flight (those data are taken from \cite{oPsTOF}).}
\label{SiO2FractionoPs}
\end{center}
\end{figure}

As reported in \cite{OurApplPhys}-\cite{cassidyPhysRevA},  systematic studies to characterize different samples that could be used as positronium converters for this experiment were performed. From those samples, we selected the so called F-samples (cf. \cite{cassidyPhysRevA}) for \ops\ production target of the experiment for the following reasons:
\begin{enumerate}
\item It provides a high production rate of $\ops$ in the range of 27-28 $\%$ for positron implantation energies between 3 and 5 keV (see Fig. \ref{SiO2FractionoPs}), thus, the high statistics required for the experiment can be reached (Section \ref{sec:sensitivity}). 
\item In the interval of implantation energies between 3 and 5 keV the fraction of \ops\ remains constant within 1$\%$ (from 28\% to 27\%), thus, the background coming from the 2$\gamma$ decay of $\pps$ remains the same for the different implantation energies. While this fraction is almost constant, the mean energy of the produced \ops\ is about a factor of 3 larger for 3 keV implantation energy that for 5 keV (see Fig. \ref{SiO2FractionoPs}). We estimated with the MC simulation that for a cylindrical cavity with 30 mm diameter and 15 mm length at 5 and 3 keV one has 0.9 and 1.9 collisions per lifetime. Therefore, the collision rate with the walls is $\sim$ 2 times bigger for 3 keV positrons and the signal is suppressed by the same factor. 
%\begin{figure}[htb!]
%\begin{center}
%\hspace{.0cm}\includegraphics[width=.8\textwidth]{inv_experiment/Target/Ncollisions.eps}
%\caption{\em  Number of collision per lifetime estimated with the MC for positron implantation energies of 3 and 5 keV.}
%\end{center}
%\label{fig:ncoll}
%\end{figure}

It is worth noting that from 3 to 5 keV the implantation depth for the positrons varies from 165 to 300 nm (the density of our porous films is about 1.5 g/cm$^3$) thus the background will not be affected because the difference of the energy absorbed in 150 nm of porous SiO$_2$ is negligible.  
\item In the samples the \ops\ is produced near-thermal energy, thus, it experiences only few collisions per lifetime with the cavity walls minimizing the suppression of the signal.
\item The target thickness is about 800 nm and it can be spin coated directly on the end plate of the beam pipe (Fig. \ref{cavity_escape}) to avoid additional material of a substrate. Therefore, the absorption of photon energy in the target will be minimized. 
\end{enumerate}

\subsection{The vacuum cavity}\label{subsec:cavity}
\begin{figure}[h!]
\begin{center}
\hspace{.0cm}\includegraphics[width=.8\textwidth]{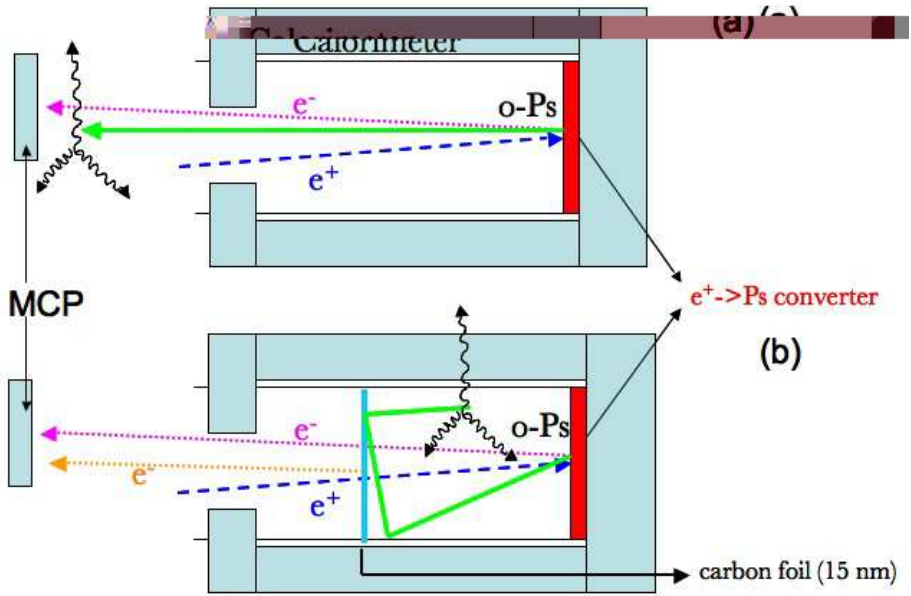}
\caption{\em (a) Ps is tagged by the secondary electrons produced by the positron that hits the converter surface (target). In the case where Ps escapes the detection region, it could mimic an invisible decay due to the strongly suppressed detection efficiency of the decay photons. (b) By closing the vacuum cavity around the $\ops$ converter one obtains a region of highly uniform detection efficiency and the \ops\ leakage through the aperture is suppressed. The 15 nm carbon foil acts as a barrier for the $\ops$ emitted from the converter but it is nearly transparent for the incoming positron beam and for the secondary electrons used for the trigger.}
\label{cavity_escape}
 \end{center}
\end{figure}

An essential issue in the experiment is that the positronium produced in the target should be confined in a region of highly uniform detection efficiency, and the \ops\ leakage through the aperture where the positrons are implanted, has to be minimized. The \ops\ which escapes the detection region, could mimic an invisible decay because in this case the detection efficiency for the annihilation photons is reduced dramatically (Fig. \ref{cavity_escape} (a)).

In their recent decay rate experiment \cite{Vallery:2003iz}, the Michigan group tried to minimize this effect to reduce the systematic error, using a double chamber-cavity, but still the escape probability for \ops\ was at 200 ppm.
In our measurement such a leakage would limit the sensitivity to a \binvdecay\ $\simeq 10^{-6}$.
Therefore, to be able to reach the aimed sensitivity of the experiment, one has to invent a new method. We plan to completely close the \ops\ formation cavity \footnote{Only microscopic holes will be left for pumping the vacuum inside the cavity, leaving a negligible probability for the \ops\ to escape the detection region.} by employing a thin (15 nm) carbon film in which the few keV positrons can pass through. However, this will block even the most energetic positronium (several eV). Furthermore, with this method an additional signature for the presence of a positron in the formation cavity is added to the trigger scheme: the coincidence between the SE from the target and the ones produced in the carbon foil (Fig. \ref{cavity_escape} (b)). Hence, the S/N ratio will be further enhanced without a dramatic loss of the trigger rate. This point is presented in Section \ref{subsec:tagging}, some preliminary measurements have been performed in order to test this idea. The carbon films that are planned to be used are currently employed at PSI\footnote{PSI, Paul Scherrer Institut, CH-5232 Villingen-PSI, Switzerland} in the muonic-hydrogen experiment \cite{toccione} for tagging the muons. Similar to our application they are used for detecting the SE emitted by the crossing muons. Some of those films were kindly given to us from the PSI group and we tested their SE emission yield in the implantation positron energy range of 1-7 keV and their permeability to \ops. The great advantage of this technique is that the \ops\ leakage is expected to be reduced to a negligible level.

% This is done comparing the TOF spectra of the SiO$_2$ films with and without the thin carbon film in front of our SiO$_2$ positron converter (in this way one can measure the SE emission yield as well). We also varying the distance of the carbon foil to the target to understand if the pick-off probability is affected.
 %Furthermore, with this method an additional signature for the presence of a positron in the formation cavity is added to the trigger scheme as we will discuss in the next section.

In order to minimize the energy loss of the annihilation photons in the vacuum cavity, one should design the beam pipe as thin as possible. In the H1 experiment at DESY, a vacuum pipe with 840 microns wall thickness was constructed. The internal part consisted of 40 microns aluminum, while the external part was made of carbon (800 microns). Similarly, we are designing the pipe for our experiment \footnote{We are also considering the possibility of making a vacuum pipe out of a scintillating material with the internal part coated with a thin aluminum foil.}.

To avoid problems with the magnetic field necessary to guide the positrons to the target region, the coil should be wound directly on the beam pipe, since we demonstrated with our TOF and PALS setup (\cite{cugthesis}), that in this configuration the gain of the PMTs is not affected. Therefore, the material necessary for the coil should also be minimized. A coil made with a copper wire of 100 microns can produce the necessary field of 60 Gauss with a current of 0.5 A (the diameter of a wire fusing at this current is about 20 microns). In the following discussions, we consider a beam pipe with 1 mm wall thickness, thus the expected sensitivity should be understood as a conservative result (Section \ref{sec:sensitivity}).

\begin{figure}[h!]
\begin{center}
\hspace{.0cm}\includegraphics[width=.8\textwidth]{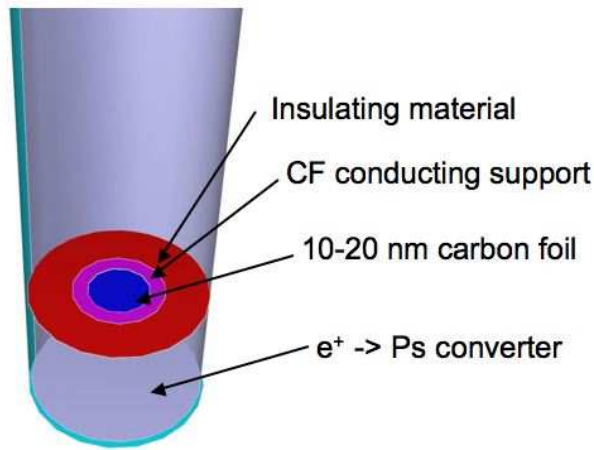}
\caption{\em Schematic of the target region and the carbon foil.}
\label{fig:CFDetails}
 \end{center}
\end{figure}

The carbon foil and its support could be suspended by some thin wires so that it is possible to move it in vacuum (see Fig. \ref{fig:CFDetails}). With such a scheme one could vary the distance between the target and the carbon foil, i.e., one would change the number of collisions that \ops\ will suffer during its lifetime. The signal will be suppressed for shorter distances in the same way as with the method described before by applying different positron implantation energies. 
%We are performing some tests in order to understand how this will affect the background. 

\subsection{Positron tagging system}\label{subsec:tagging}
\begin{figure}[h!]
\begin{center}
\hspace{.0cm}\includegraphics[width=.8\textwidth]{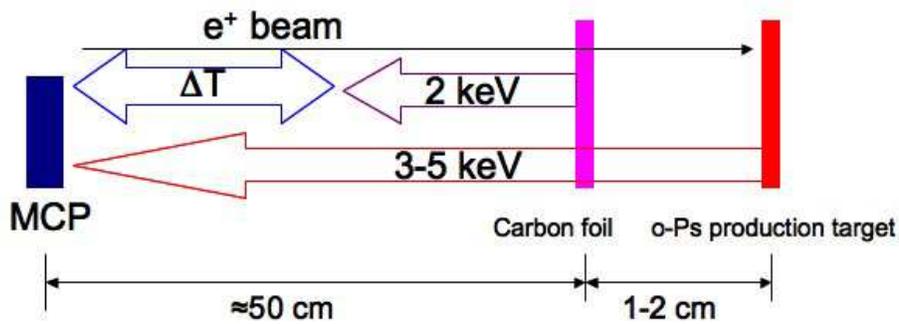}
\caption{\em Schematic of the positron tagging system for the experiment using the MCP signal for secondary electrons emitted from the target and from the thin carbon foil.}
\label{tagging}
 \end{center}
\end{figure}

The positron tagging system (see Fig \ref{tagging}) is based on the high performance MCP (Hamamatsu F4655-12) as a SE detector (noise rate $<$ 1Hz).
The coincidence between the SEs produced by positrons hitting the carbon foil and the SEs emerging from the \ops\ production target will be used to tag the positron appearance in the target region.

\begin{figure}[h!]
\begin{center}
\hspace{.0cm}\includegraphics[width=.8\textwidth]{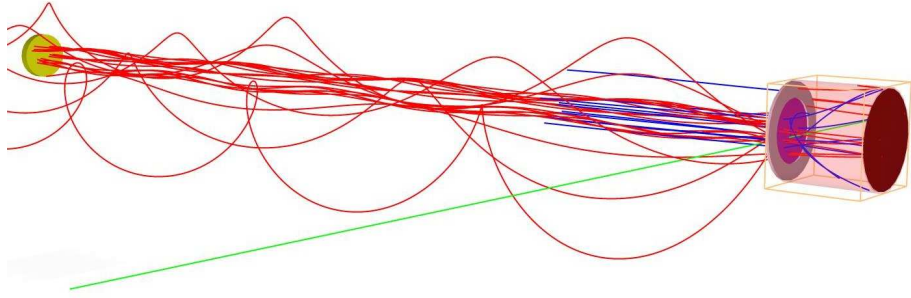}
\caption{\em Trajectories of the positron (blue) and secondary electrons (red) for the new design with the carbon foil.}
\label{trajectories}
 \end{center}
\end{figure}

\begin{figure}[h!]
\begin{center}
\hspace{.0cm}\includegraphics[width=.8\textwidth]{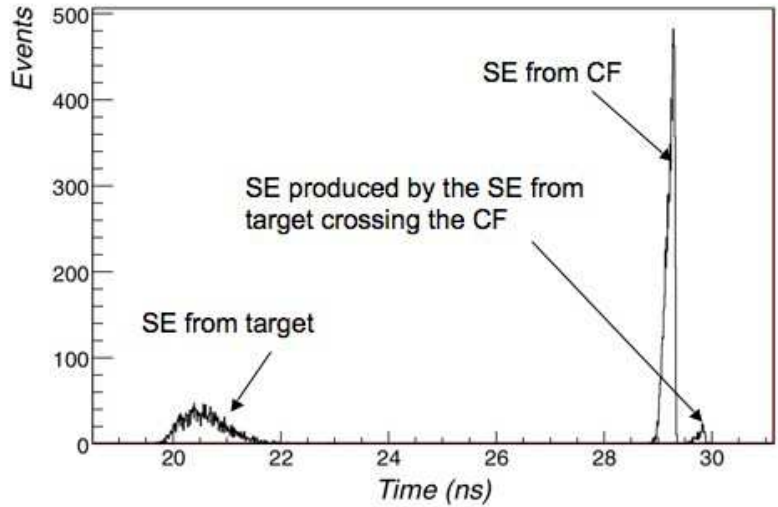}
\caption{\em Simulated time distributions of the secondary electrons detected in the MCP. See text for details. }
\label{MCPTaggingSignal}
 \end{center}
\end{figure}

The acceleration voltage for the positrons is applied in two steps. The positrons are first accelerated to 2 keV energy by the voltage applied between the ground pipe and the carbon foil, thus the SEs emerging from the carbon foil (CF) are transported with this energy back to the MCP. In the second acceleration stage, the positrons are accelerated by a 1-3 keV voltage applied between the CF and the target. Once they hit the target, the SEs are transported back to the carbon foil and the electrons able to cross it are then additionally accelerated so that their final energy will be 3-5 keV (see Figure \ref{trajectories}). Figure \ref{MCPTaggingSignal} shows a MC simulation of the time distribution of the MCP signals from different SE sources; the time zero point is defined by the incoming beam positron crossing the carbon foil. The two prominent peaks are due to the SEs emitted when the positrons hit the target (at about 21 ns) and the SEs from the carbon foil (at about 29 ns) when a positron crosses the foil. The SEs from the target are accelerated with a higher voltage than those from the foil and reach the MCP first, although they start later and have a larger flight path. The small third peak visible in the time spectrum of the MCP is  produced by additional secondary electrons generated by some SEs from the target interacting in the carbon foil. Those will be accelerated by the same voltage that accelerates the positrons and, therefore, they are detected 1-1.5 ns after the second peak (the cavity length used in the simulation was 30 mm and the positron energy is 6 keV). The broadening of the first peak is due to the angular spread of the SEs emitted from the target and scattering in the carbon foil. 
%It is important to apply in this first stage the same voltage for both measurement with 3 and 5 keV in order to have the same number of positrons annihilations in the carbon foil, i.e. to have the same background.
 %The two voltage values and the MCP position were optimized with the help of the simulation to get a good separation of the two MCPs signal coming from the SE emitted from the target and SE emitted from the carbon foil. 

 %The expected efficiency of tagging a positron with this technique will be about 20-10 $\%$ varying the voltage applied to the target from 3 to 5 keV. In fact, the number of secondary electron emitted depends on the implantation energy of the positrons.

\subsubsection{Trigger efficiency and confidence level} \label{sec:Trigger_CL}

\begin{figure}
\begin{center}
%%\hspace{2.cm}\includegraphics[width=0.8\textwidth]{opsdecayrate.eps}
\includegraphics[width=0.7\textwidth]{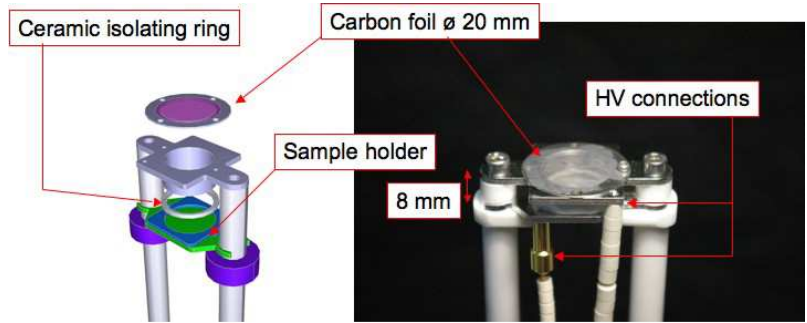}
\caption{\em Setup used for the measurements with the tagging system. In order to accelerate the SEs from the target a voltage difference between the sample holder (green) and the carbon foil has to be applied. A ceramic ring electrically insulates the carbon foil holder and the target holder.}
\label{trigger_ass}
\end{center}
\end{figure}

In order to estimate the trigger efficiency and fake trigger suppression of the tagging system described above, a test set-up has been constructed (see Fig. \ref{trigger_ass}). In this set-up the distance between the carbon foil and the target was 8 mm and the distance separating the carbon foil and the MCP was 90 cm.  

\begin{figure}[h!]
\begin{center}
%%\hspace{2.cm}\includegraphics[width=0.8\textwidth]{opsdecayrate.eps}
\includegraphics[width=0.7\textwidth]{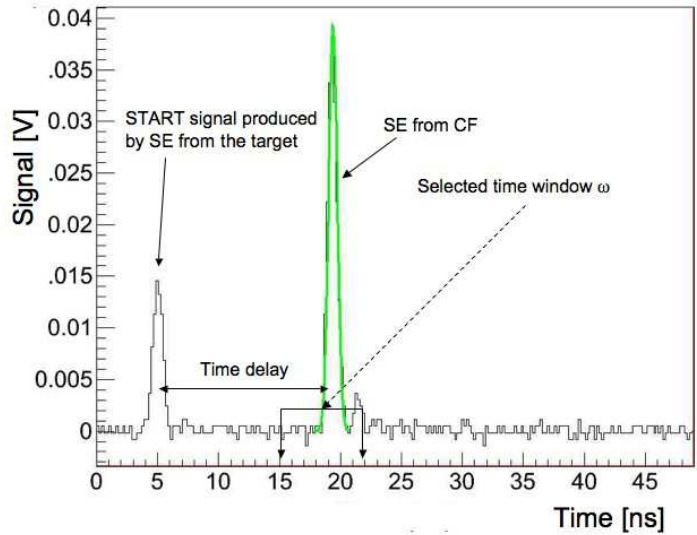}
\caption{\em Example of an event acquired with an oscilloscope. The first signal is produced by the SE from the target and defines the START signal. The second pulse is produced by the SE from the CF $\sim 15 ns$ after the START signal. The time window $\omega$ between 15 and 22 ns (i.e. between 10 and 17 ns delay from the START signal) is defined in order to calculate the trigger efficiency and the fake trigger ratio. A Gaussian fit is used in order to determine the position of the pulses (green). The origin of the third peak are the SE produced by the SE from the target crossing the CF as expected from the simulation (see Fig. 13).}
\label{ScopeSignal}
\end{center}
\end{figure}

\begin{figure}[h!]
\begin{center}
%%\hspace{2.cm}\includegraphics[width=0.8\textwidth]{opsdecayrate.eps}
\includegraphics[width=0.7\textwidth]{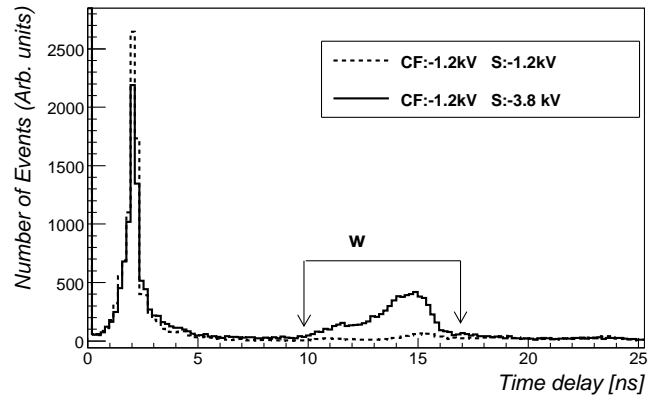}
\caption{\em Measured delay of the SE from the carbon foil relative to SE from the START signal. The dashed line represents the data obtained by applying the same voltage on the carbon foil and on the sample (-1.2kV). The solid line is obtained by applying -1.2 kV on the carbon foil and -3.8kV on the sample. In this case the broadening of the peak at around 15 ns is mainly due to variations of the flight time of the SE from the target that produce the START signal. The peak at about 2 ns are due to the MCP deadtime (see text for details). }
\label{SE_delay_10mV}
\end{center}
\end{figure}

The measurements were performed by acquiring the MCP timing with a 5 Giga sampling oscilloscope (Lecroy wave-runner 44XMi). The threshold for the acquisition was set to 10 mV and a sample of $10^{5}$ events was recorded in a time window of 50 ns after the arrival time at the MCP of the first electron that was defined as START signal. The start signal can be produced by the SEs from the carbon foil or by the SE produced at the target. The probability to detect a SE coming from the target is suppressed by the angular spread introduced by crossing the carbon foil, when propagating in the backward direction. This introduces in the trajectories larger spirals along the magnetic field axis with respect to the SE from the CF, and thus, reduces the probability of the SE to hit the MCP active area. An example of an event recorded with the oscilloscope is shown in Fig. \ref{ScopeSignal}. The START signal is given by SEs produced at the target and the pulse at 20 ns is from the SEs produced at the CF. The data in Fig. \ref{SE_delay_10mV}  show the time delay spectrum with respect to the START signal for the electrons arriving at the MCP with $10^{5}$ triggers. These results confirm qualitatively the prediction of the simulation\footnote{As will be discussed later in this section, Geant 4 does not reproduce correctly the transmission, scattering angles and backscattering coefficients of charged particles with few keV energies.}. Note, that in this distribution the time is inverted with respect to the simulations. In the simulation the time t=0 is defined by the positron arrival time at the CF.

The first peak at about 2ns (and a tail extending to 5 ns) is due to the MCP dead time: the first electron in the SE cloud from the target (or carbon foil) makes a START signal and after the dead time a second electron of the same cloud produces another signal. The second peak between 10 and 17 ns is triggered by the SEs from the CF when the START signal is produced by SE from the target; only the events in this peak are used for the positron tagging. The broadening of this peak is due to the angular spread of the SEs emitted from the target and scattering in the carbon foil.
%This effect was qualitatively predicted by the MC as we mentioned before. 

%The difference of the delay between the measurements with stainless steel and silica sample will be discussed later in this chapter.

The time separation of 15 ns between the START and the second peak is consistent with a simple estimation considering a straight propagation of the SE electrons produced at the carbon foil and the SEs from the sample (see Table \ref{table:SE_flight}). 

\begin{table}[h!]
\begin{center}
{\begin{tabular}{lllc}
\hline
%\\
\vspace{-0.1cm}
Voltage [kV] & Flight time [ns] &  \\
\hline
1.2 & 39 &  \\
3.8 & 22 & $\Delta t$ 17 ns\\
\hline
\end{tabular}}
\caption{\em Estimated flight time for SE emerging from CF (1.2 kV) and from the sample 3.8 kV for a distance of 80 cm. The expected delay time is 17 ns.  }
\label{table:SE_flight}
\end{center}
\end{table}

To estimate the trigger efficiency and the fake trigger suppression of our system, we compare two measurements: 
\begin{itemize}
\item  a set of data with a voltage difference between the carbon foil and the sample of -1.2 kV and -3.8 kV, respectively (solid line in Fig. \ref{SE_delay_10mV}),
\item  a set of data with the same voltage applied on the carbon foil and the sample (-1.2 kV) (dashed line in Fig. \ref{SE_delay_10mV}). 
\end{itemize}

We define the trigger efficiency $\epsilon_{SE}$ as:
\begin{equation}\label{eq:trigg_eff}
\epsilon_{SE}= \frac{N_{w}}{N_{tot}}
\end{equation}  
where $N_{w}$ is the number of events arriving with a time delay from the start signal in a selected time window $w$ when the a voltage difference is applied between the target and the carbon foil. $N_{tot}$ is the total number of START signals. 

The carbon foil suppression factor $\kappa_{SE}$ is defined as:
\begin{equation}\label{eq:trigg_fake}
\kappa_{SE}= \frac{N^{'}_{w}}{N_{w}}
\end{equation}  
where $N^{'}_{w}$ is the number of events in the selected time window $w$ when the same voltage is applied to the sample and the CF (i.e. SE produced at the target cannot reach the MCP). In this case the probability for SEs emitted from the sample to reach the MCP is strongly suppressed because the transparency of the foil for those electrons is practically zero.  The START signal and the peak at $\sim$ 2 ns for the dashed line in Fig.  \ref{SE_delay_10mV} (the same voltage on target and CF) is produced by SEs from the carbon foil, emitted by an incoming positron, since (almost) no SE from the target reach the MCP. Thus, $\kappa_{SE}$ gives suppression of fake triggers obtained by the coincidence positron trigger requirement for events with a positron traversing the carbon foil, but no SE emitted from the target. It is used to estimate the background from positrons, which do not reach the target, e.g. due to backscattering from the carbon foil. This background will be discussed in Sections \ref{sec:backPsfoil} and \ref{sec:backPositrons}. Note, that the value $\kappa_{SE}$ (see below) is not in contradiction with the requirement for the invisible decay search experiment to have a fake trigger fraction with no energy deposition in the ECAL of $< 10^{-7}$. The suppression $\kappa_{SE}$ refers to events with a positron passing the carbon foil and for most of these events the annihilation photons will deposit energy in the ECAL.

%The peak in Fig. \ref{SE_delay_10mV} results from the most energetic SEs that are able to cross the CF, thus our estimation is very conservative. The suppression factor $\kappa_{SE}$ represents the ratio of fake triggers related to positrons that are not accompanied by SE from the target and gives an estimation of the suppression of the backgrounds generating to positrons that does not enter in the region between the carbon foil and the target. These backgrounds will be discussed later in Section \ref{sec:backPsfoil} and Section \ref{sec:backPsfoil}. Note, that the value extrapolated the carbon foil suppression factor is not in contradiction with fake positron triggers fraction of $10^{-7}$ required by the experiment. In fact $\kappa_{SE}$ , is referred to the confidence of having a positron in the vacuum cavity by using the SE coming from the target in coincidence with the SE from the CF. Thus, this value is not related to fake triggers when a positron is not in the vacuum cavity.

The signals of SEs emerging from the sample and those emerging from the CF can be further discriminated by their charge. The large scattering angle acquired by the SEs from the target crossing the CF drastically reduces the probability to have, as a trigger, a signal produced by more than one electron. On the other hand, the STOP signal is composed of a larger number of SE because it is produced by the sum of:
\begin{itemize}
\item  the SE emitted by positrons crossing the CF;
\item the SE that are emitted when, the SE produced at the target, cross the carbon foil in the opposite direction, since the small difference in their arrival time at the MCP (relative to the SE from the oncoming positron) is not resolved because of the small distance (8 mm) between the target and the carbon foil and the large voltage between them.
\end{itemize}

\begin{figure}[h!]
\begin{center}
%%\hspace{2.cm}\includegraphics[width=0.8\textwidth]{opsdecayrate.eps}
\includegraphics[width=0.6\textwidth]{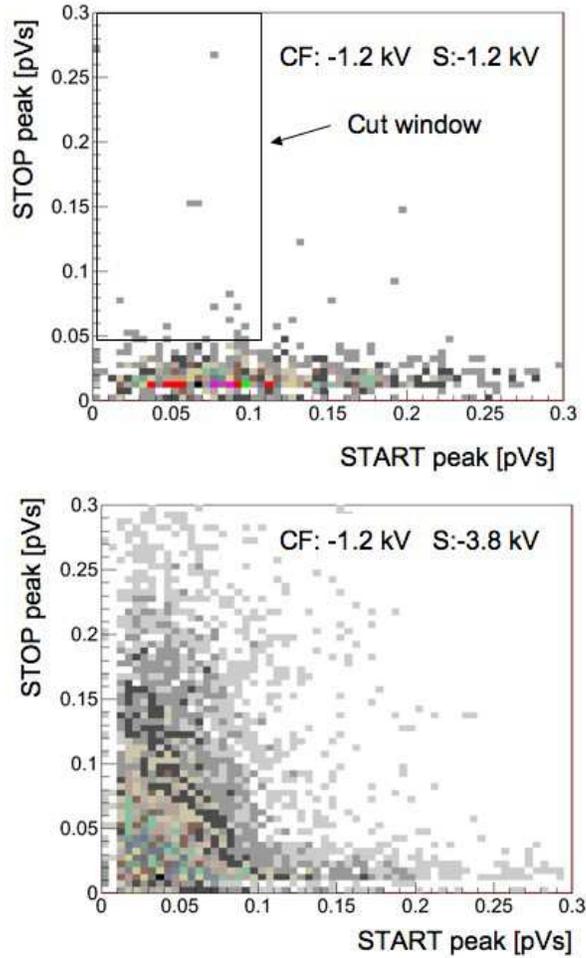}
\caption{\em Charge (measured with 50 $\Omega$ termination) in pVs of the start signal versus the charge of the signal for events in the time window 12-17 ns. Top: the same voltage of 1.2 keV is applied to the target and the carbon foil. Bottom: a voltage of 3.8 keV is applied to the target and 1.2 keV to the carbon foil.}
\label{E_corr}
\end{center}
\end{figure}

\begin{figure}[h!]
\begin{center}
\includegraphics[width=0.6\textwidth]{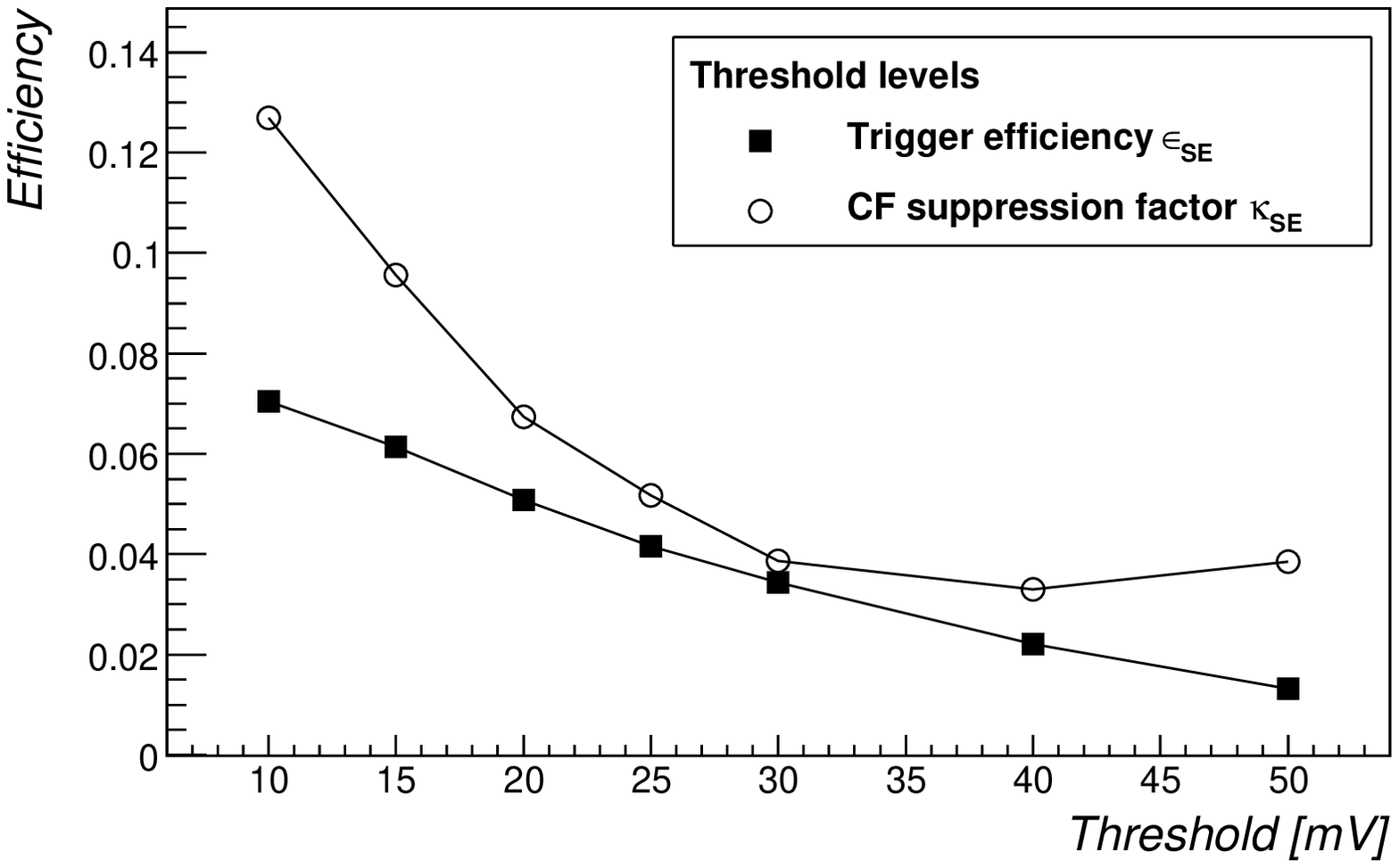}
\includegraphics[width=0.6\textwidth]{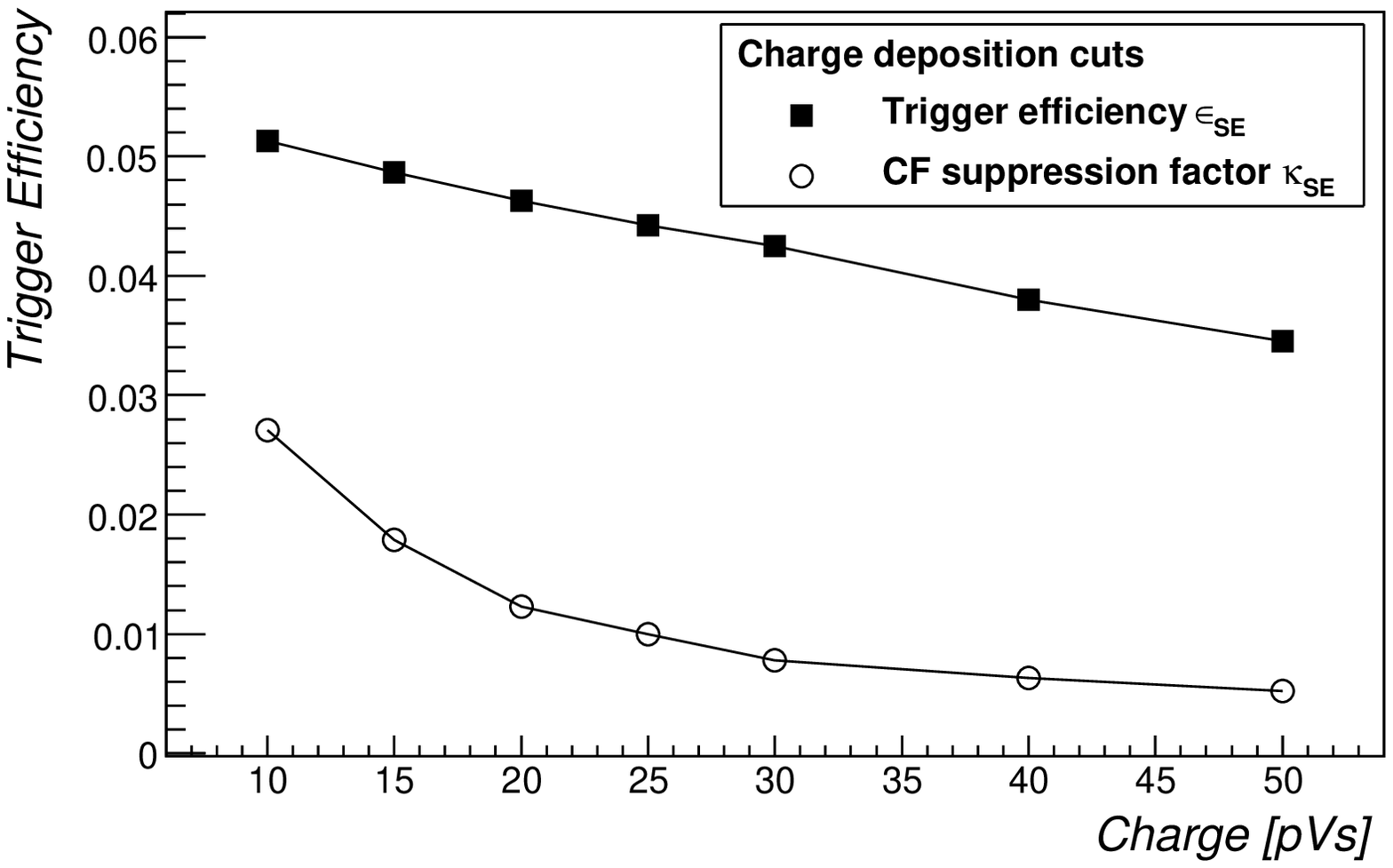}
\caption{\em (Top plot) Trigger efficiency (boxes) and coincidence suppression factor $\kappa_{SE}$ (circles) as a function of the threshold set on the signal in the time window 12-17 ns. (Bottom plot) Trigger efficiency (boxes) and  coincidence suppression factor $\kappa_{SE}$ (circles) as a function of the cut on the charge (voltage integral) on the signals in the time window 12-17 ns.}
\label{E_cut_eff}
\end{center}
\end{figure}

This is clearly visible in the Fig. \ref{E_corr} where the charge of the START signal (i.e. the first electrons which hits the MCP) is plotted versus the signal in the time window of 12-17 ns for both voltage configurations. This effect can be exploited to considerably increase the confidence level for a positron in the vacuum cavity without dramatically affecting the trigger efficiency as shown in Fig. \ref{E_cut_eff}. The trigger efficiency (boxes) and the coincidence suppression factor $\kappa_{SE}$ (circles) are studied for two different selection cuts on the MCP signal:
\begin{enumerate}
\item by applying a threshold on the MCP stop signal;
\item by applying a cut on the charge (integral over the signal) produced by the MCP.
\end{enumerate}
In the first case the fake trigger ratio cannot be suppressed below a value of 4$\times 10^{-2}$ while the trigger efficiency drops to a value of $10^{-2}$. In the second case, the cut performed on the charge shows to be much more efficient. For a cut on the charge of 40 pVs is suppressed to 5$\times 10^{-3}$ with a corresponding trigger efficiency of 4$\times 10^{-2}$.

% An essential issue is the production of $\ops$ at the carbon foil surface. In fact, the confidence level of the $\ops$ tagging using the coincidence of the SEs from the carbon foil and the sample is not $100 \%$. 

%\begin{figure}
%\begin{center}
%\includegraphics[width=0.7\textwidth]{ESTAR_G4.eps}
%\caption{\em Simulation of the SEs flight to the MCP. The carbon foil introduces a scattering angle for both incoming positrons and crossing SEs suppressing the trigger efficiency.}
%\label{SE_transport}
%\end{center}
%\end{figure}

%\begin{figure}
%\begin{center}
%\includegraphics[width=0.7\textwidth]{Scattering_2-3-4kV.eps}
%\caption{\em Simulation of the SEs flight to the MCP. The carbon foil introduces a scattering angle for both incoming positrons and crossing SEs suppressing the trigger efficiency.}
%\label{SE_transport}
%\end{center}
%\end{figure}

\subsection{The photon detector}\label{subsec:InvDet}

\begin{figure}[h!]
\begin{center}
\hspace{.0cm}\includegraphics[width=.7\textwidth]{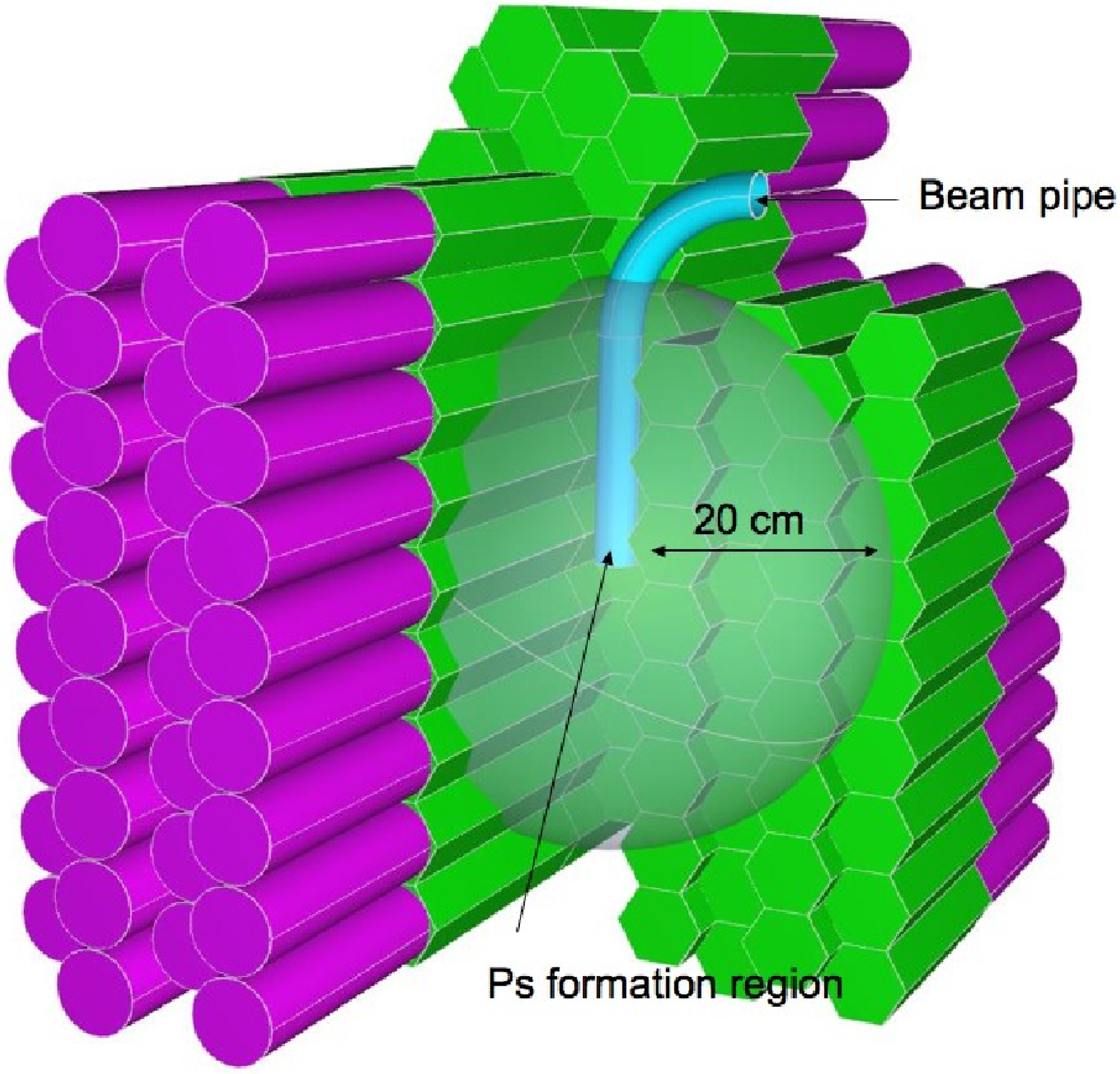}
\caption{\em  Cross section of the BGO calorimeter mounted around the beam pipe. The sphere is to show that the minimal BGO thickness around the target is of the order of 200 mm.}
\label{DMDet}
\end{center}
\end{figure}

The same BGO crystals as in our former \ops\ invisible decay search will be used to detect $\gamma$-quanta produced in positron or positronium annihilation. The geometry of the detector should be modified to accommodate the beam pipe as we proposed in \cite{Gninenko:2004ft} (see Fig. \ref{DMDet}). The test of the trigger scheme presented in Section \ref{subsec:tagging} was done in a straight geometry, however the end of the beam pipe should be bent in order to avoid a direct way for the annihilation's photons and for Ps to escape the detection region. We are not expecting a decrease of the trigger efficiency or an increase in the background because our simulation predicts that with a careful design the transportation efficiency for both positrons and SE through the curved beam pipe region is 100\%.  
 Furthermore, the crystals have to be arranged perpendicularly to the B-field, since, as we confirmed in our TOF and PALS setup (\cite{cugthesis}), in this configuration the magnetic field will not affect the performance of the PMTs.

The $\gamma$-detector serves to veto effectively the positron annihilations into photons. It has been shown that 
its inefficiency for the detection of annihilation's photons is less then $10^{-9}$ for the energy threshold of $\simeq 80 $ keV 
\cite{oPsInv,thesis}\footnote{It is worth noting that in the experiment described in this section the threshold will be lower because the pileup will be reduced by a factor 20.}.
Moreover, we realized that the energy resolution of the crystals could be increased by replacing the Teflon in which the crystals are wrapped with the 3M radiant foil. This reduces the amount of dead material by a factor 12. Furthermore, all the BGO crystals surface is now polished. By roughing the BGO surface except where the PMT is coupled to the crystal, the energy resolution of the calorimeter could be further improved.
Therefore, the calorimeter could be refurbished in order to increase the energy resolution resulting in an improvement of the detector performance. 

\section{Background estimation}\label{sec:Bkg_vac}

As mentioned in the previous section, we expect backgrounds which originate from the following sources:
\begin{enumerate}
\item the annihilation energy losses that are estimated to be at a level of $\simeq 10^{-7}$. 
\item a fake positron tagging at a level $\simeq 10^{-7}$
\end{enumerate}

\begin{figure}[h!]
\begin{center}
\includegraphics[width=0.6\textwidth]{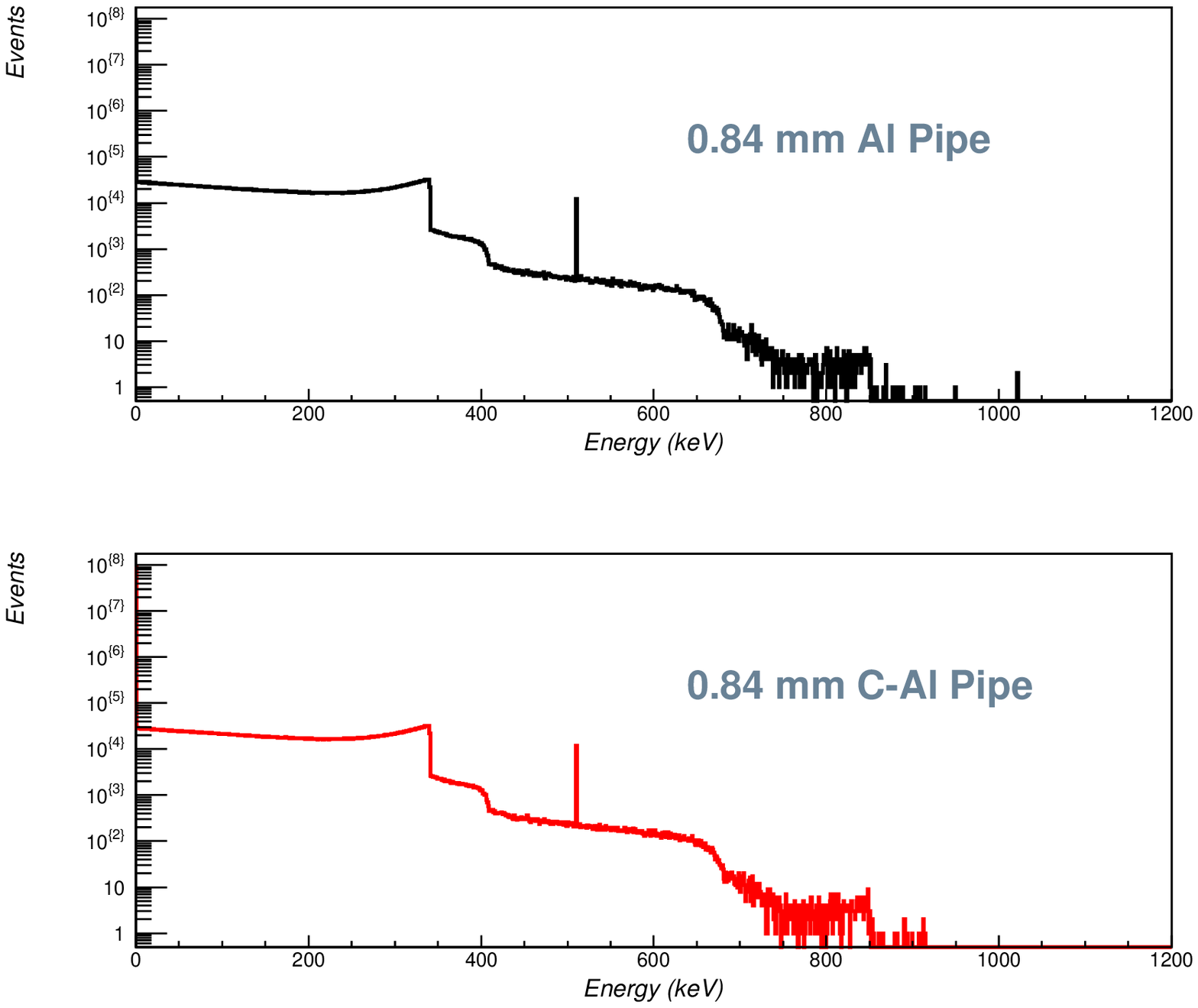}
\caption{\em Distributions of the energy deposited in the dead material surrounding the target region (target substrate, beam pipe and copper coil) from annihilation events in the target. The top plot shows the results of the MC simulation for a 0.84 mm thick aluminum pipe. The bottom plot the distribution for the pipe construction with 0.04 mm aluminum and 0.800 mm carbon. The total number of simulated $2\gamma$-events is $10^8$ in both cases. The peaks at 511 keV and 1022 keV correspond to the total photo-absorption either of a single 511 keV photon or of both of them, respectively.}
\label{beampipesim}
\end{center}
\end{figure}

The main contribution to the first background of the list above is coming from the thickness of the vacuum beam pipe. In Fig. \ref{beampipesim}, we show the simulation results of the energy deposited in a 0.84 mm thick aluminum pipe and a pipe with 0.04 mm aluminum and 0.800 mm thick carbon pipe (similar to the one that was used at the H1 experiment at DESY). In these distributions, the energy deposited in the target substrate and in the copper wire surrounding the beam pipe are also included. 
%  As for the design of the previous experiment (see Section \ref{}), the MC simulation reveals to be an essential tool to identify and understand how to suppress these backgrounds.

Different possibilities of fake triggers that can be produced in this experiment have been identified:

\begin{itemize}
\item Fast backscattered $\ops$ produced at the carbon foil surface have a certain probability to escape the detection region. 
\item Backscattered positrons from the carbon foil surface or the target have a probability to escape the detection region. 
\end{itemize}

\subsection{Fast backscattered $\ops$ from the carbon foil} \label{sec:backPsfoil}

\begin{figure}[h!]
\begin{center}
%%\hspace{2.cm}\includegraphics[width=0.8\textwidth]{opsdecayrate.eps}
\includegraphics[width=0.6\textwidth]{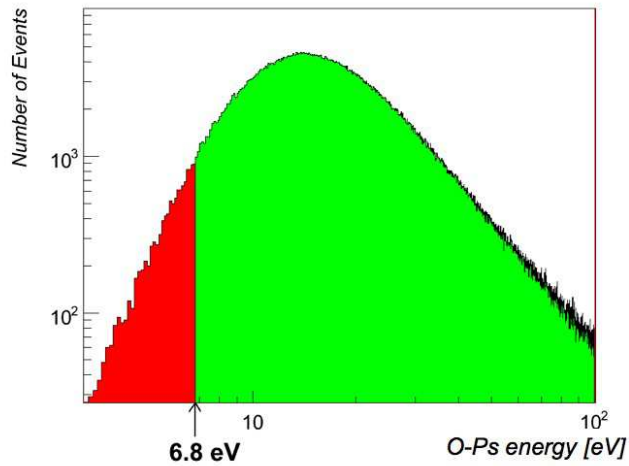}
\caption{\em Energy distribution of fast \ops\ originating from backscattered $e^{+}$. The pick-off probability after a collision of \ops\ with the walls is taken to be zero for \ops\ energies smaller than the 6.8 eV while for higher energies it is 0.95.}
\label{Fast_ops_distr}
\end{center}
\end{figure}

\begin{figure}[h!]
\begin{center}
%%\hspace{2.cm}\includegraphics[width=0.8\textwidth]{opsdecayrate.eps}
\includegraphics[width=0.9\textwidth]{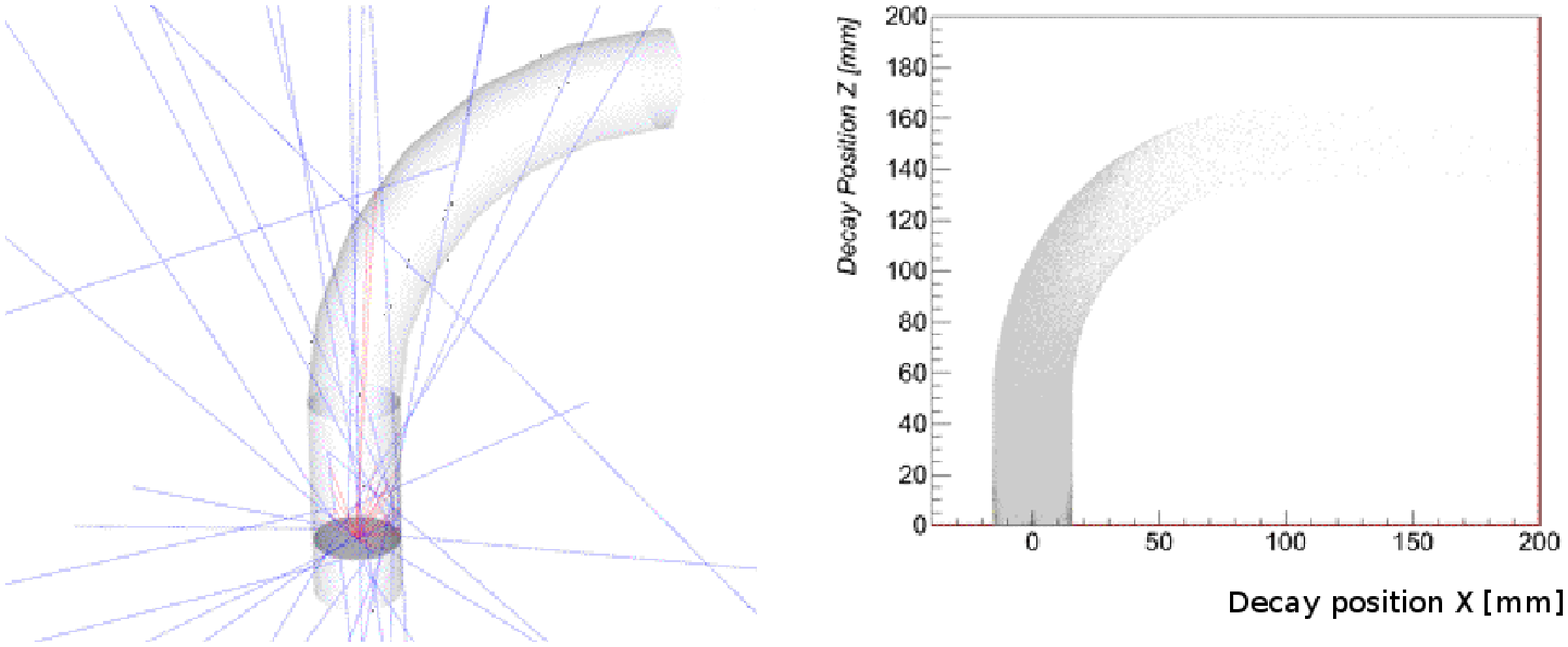}
\caption{\em Simulated decay position of \ops\ in the vacuum cavity. The escaping probability estimated with the MC simulation is $\simeq 1\times10^{-4}$.}
\label{Fast_ops_esc}
\end{center}
\end{figure}

Fast backscattered $\ops$ can be produced from shallow implanted positrons that capture an electron exiting the surface. The contribution to the background arising from this effect has been studied with the MC simulation. We approximated the energy distribution of $\ops$ with a Landau distribution peaking around 15 eV \cite{howell} (see Fig. \ref{Fast_ops_distr}). The probability $P(E)$ of $\ops$ to annihilate via pick-off when it collides with the cavity walls was also included. As a function of the energy, $P(E)$ can be divided in two regions (\cite{gidley1995}): 
\begin{enumerate}
\item For $E_{\ops}<6.8 eV$ ,  $P(E)=0$
\item For $E_{\ops}>6.8 eV$ ,  $P(E)=0.95$
\end{enumerate}
Both assumptions are conservative. In fact, even $\ops$ with an energy smaller than its binding energy (6.8 eV) can undergo pick-off annihilation, thus the number of Ps atoms escaping the detection region will be less than what we estimated with the simulation using the assumption 1). The same is true for Ps with energies above its bounding energy because in this case the probability to dissociate in a collision is close to 100\% \cite{gidley1995}. We define the escaping probability $\xi$ as:
\begin{equation}\label{eq:trigg_eff}
\xi_{\ops}= \frac{N_{escape}}{N_{tot}}
\end{equation}  
where  $N_{escape}$ is the number of $\ops$ which decay after the bending of the vacuum cavity (See Fig. \ref{Fast_ops_esc}) and $N_{tot}$ is the total number of events simulated. Note that this approach is very conservative because the annihilation gammas produced outside the bending region have a non-zero probability to deposit some energy in the calorimeter. The escaping probability estimated with the simulation (Fig. \ref{Fast_ops_esc}) is $\xi_{\ops} \simeq 1 \times 10^{-4}$. In order to calculate how this value affects the sensitivity of the experiment, it has to be multiplied with the fast $\ops$ formation probability \cite{howell} ($<10\%$). In order to mimic an invisible decay, the fast $\ops$ escaping event has to coincide with an accidental trigger. Thus, the carbon foil suppression factor $\kappa_{SE}$ presented in Section \ref{sec:Trigger_CL} further suppresses this background to a level  $< 5 \times 10^{-8}$.

\subsection{Backscattered positrons}\label{sec:backPositrons}
\begin{figure}[h!]
\begin{center}
%%\hspace{2.cm}\includegraphics[width=0.8\textwidth]{opsdecayrate.eps}
\includegraphics[width=0.6\textwidth]{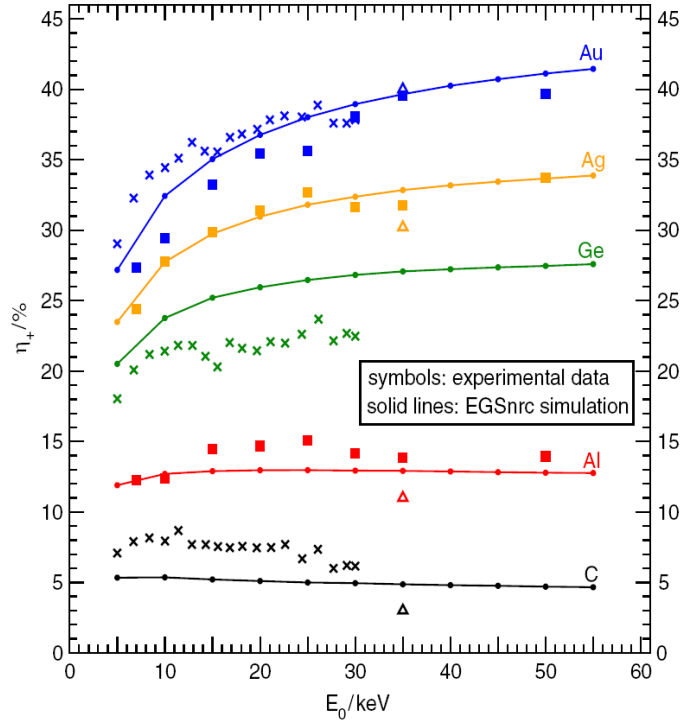}
\caption{\em Positron backscattering coefficients ($\eta_{+}$) versus positron kinetic energy \cite{EGSnrc}.}
\label{BackEGSnrc}
\end{center}
\end{figure}

 The positrons are transported along the beam line with an energy that can be varied from $E_{0}=10-200$ eV. The carbon foil and the target are biased with a potential with respect to the beam pipe, giving an additional acceleration to the positron, thus, their energy at the carbon foil is $E_{TOT}=E_{0}+E_{CF}$ and at the target it is $E_{TOT}=E_{0}+E_{Target}$. If a positron backscatters in the carbon foil (or in the target) it is decelerated by the applied voltage. In this case, it can either be re-implanted or escape the detection region. 
The fate of the backscattered positron depends on its energy loss in the backscattering process at the foil or target $\Delta E = E_{TOT}-E_{Back}$: 
\begin{enumerate}
\item if $E_{Back}  <E_{CF(Target)}$  the positron is re-implanted in the carbon foil (or in the target) with an energy $E_{Back}$. 
\item if $E_{Back}  > E_{CF(Target)}$ the positron has enough energy to escape the carbon foil (the target) voltage and it can be transported back in the beam line with an energy $E_{Back}-E_{CF(Target)}$
\end{enumerate}
This second process represents a dangerous source of background. An escaping backscattered positron could give a trigger and be transported back by the magnetic field outside the detection region, and thus, it will not deposit any energy in the calorimeter. This effect can be suppressed by decreasing the beam transport energy $E_{0}$ so that the minimum energy loss necessary to escape from the target voltage is smaller and consequently  also the escaping probability is suppressed. In this section, a precise estimation of this effect will be carried out using MC calculations.  The dominant process in the backscattering effect is the positron multiple-scattering. GEANT4 has a set of low energy classes that can be included in the Physics list using PENELOPE MC cross sections such as {\it G4PenelopeBrehmsstrahlung, G4PenelopeComptonEffect}. Unfortunately, the multiple-scattering process is not yet implemented and consequently it is not possible to estimate the backscattering coefficients with GEANT4. Accurate knowledge of the energy and angular distributions of keV charged particles is possible with the EGSnrc (Electron Gamma Shower) Monte Carlo code \cite{EGSnrc}. This code includes sophisticated low-energy physics comparable with PENELOPE. Moreover, EGSnrc is considerably faster than PENELPOPE and does not require careful tuning of the simulation parameters. In EGSnrc a special package was developed for the calculation of backscattering coefficients. The performance of EGSnrc for keV backscattering particles is presented in \cite{EGSnrc} where MC simulation data are compared to experimental results for both electron and positron backscattering (see Fig. \ref{BackEGSnrc}).  \\
We define the positron escaping ratio $\epsilon_{e^{+}-esc}$  from the detection region as:
\begin{equation}\label{eq:escape_back_positron}
\epsilon_{e^{+}-esc}= N_{E_{II}}/N_{tot}
\end{equation}
where $N_{E_{II}}$ is the number of backscattered positrons whose longitudinal energy loss is smaller than the incoming positron energy before the acceleration.

The effect of the positron backscattering has been studied in two situations:
\begin{enumerate}
\item Positron backscattering at the carbon foil. 
\item Positron backscattering at the $SiO_2$ target. 
\end{enumerate}

\subsubsection{Positron backscattering from the carbon foil} \label{sec:backpositronfoil}
\begin{figure}[h!]
\begin{center}
%%\hspace{2.cm}\includegraphics[width=0.8\textwidth]{opsdecayrate.eps}
\includegraphics[width=0.6\textwidth]{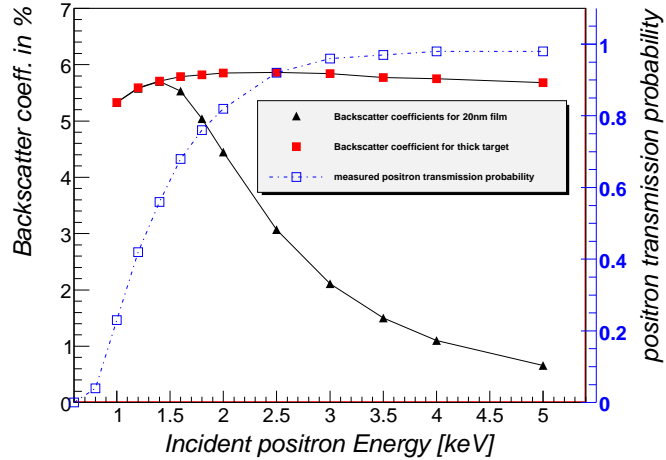}
\caption{\em Positron backscatter coefficient  ($\eta_{+}$) from a thick carbon target (black squares) and a thin carbon foil (triangles) as a function of the positron kinetic energy. The transmission probability, through the thin foil is also shown (open squares, right scale).}
\label{backcoeffCF}
\end{center}
\end{figure}

To calculate the backscattering probability,  we assume that the carbon foil has a density of $2 g / cm^{3}$ and a thickness of 20 nm. There is a substantial difference between the backscattering coefficients for semi-infinite targets and thin foils because of the transparency effect that shows a considerable suppression of the backscattering probability when the incident energy of the positron increases.  In Fig. \ref{backcoeffCF}, the backscattering coefficient $\eta_{+}$ for a thick carbon target and for a 20 nm carbon foil are compared. On the same plot the transparency of the foil for the same implantation voltages is shown.\\
The probability of a positron to escape the carbon foil voltage depends on: 
\begin{itemize}
\item the backscattering coefficient $\eta_{+}$
\item the initial positron energy $E_{0}$  before the foil acceleration $E_{CF}$
\item the backscattering energy and angular distributions
\end{itemize}
%As shown in \cite{EGSnrc}, EGSnrc allows also to calculate backscattering coefficients for thin films.

\begin{figure}[h!]
\begin{center}
%%\hspace{2.cm}\includegraphics[width=0.8\textwidth]{opsdecayrate.eps}
\includegraphics[width=0.8\textwidth]{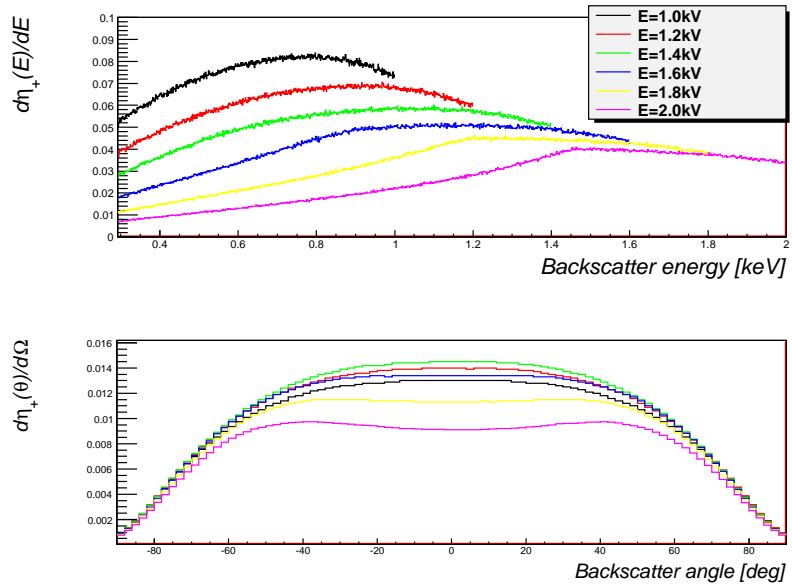}
\caption{\em (Top plot) simulation using EGSnrc. Energy spectrum of positrons backscattered from a 20 nm carbon foil for positron implantation energies in the 1-2 keV range. The cut-off energy is 300 eV.  (Bottom plot) angular distribution for backscattered positrons}
\label{E_cut_effCF}
\end{center}
\end{figure}

\begin{figure}[h!]
\begin{center}
%%\hspace{2.cm}\includegraphics[width=0.8\textwidth]{opsdecayrate.eps}
\includegraphics[width=0.8\textwidth]{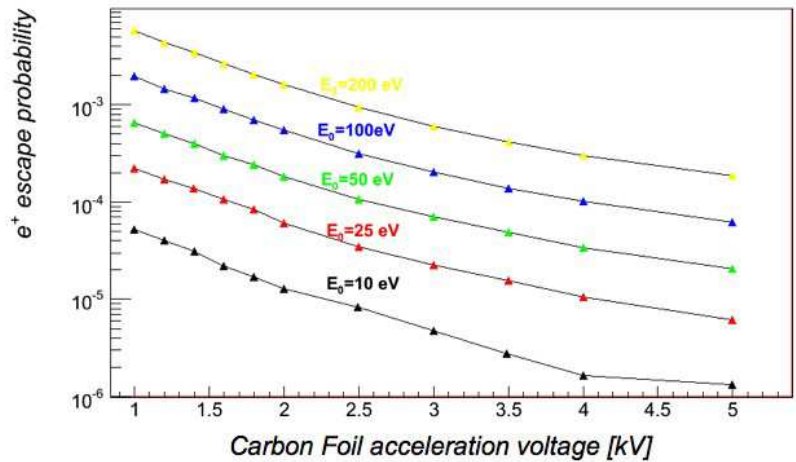}
%escape10E8_labels1-5.eps
\caption{\em Escaping probability for positrons backscattered from the carbon foil as a function of the carbon foil voltage for different positron initial energies $E_{0}$.}
\label{BKGLevelBKSCPosCF}
\end{center}
\end{figure}

In order to estimate the escaping fraction, the whole energy distribution of the backscattered positrons has to be known (see Fig. \ref{E_cut_effCF}). Figure \ref{BKGLevelBKSCPosCF} shows the escaping probability  for positrons backscattered from the carbon foil as a function of the acceleration voltage. The calculations have been performed for different positron initial energies $E_{0}$. As one can see, for a good choice of the voltage applied to the carbon foil (about 1.5 kV) and of the initial energy (the lower the better, $E_{0}=10$ eV) the escaping probability is smaller than 2$\times 10^{-5}$. Positrons backscattering at the carbon foil are not a direct source of background. To result in an artificial invisible decay, such an event has to coincide with an accidental trigger. Therefore, this background is suppressed by the confidence level of the tagging system presented in Section \ref{sec:Trigger_CL} and results in a background $< 10^{-7}$. 
%he cut-off energy of the program is 300 eV. In \cite{}, the limitation of the MC for the sub 1-kV results are discussed concluding .... {\bf ADD CONCLUSIONS OR REMOVE THIS PART} .

\subsubsection{Positron backscattering from the $SiO_2$ target}
\begin{figure}[h!]
\begin{center}
%%\hspace{2.cm}\includegraphics[width=0.8\textwidth]{opsdecayrate.eps}
\includegraphics[width=0.8\textwidth]{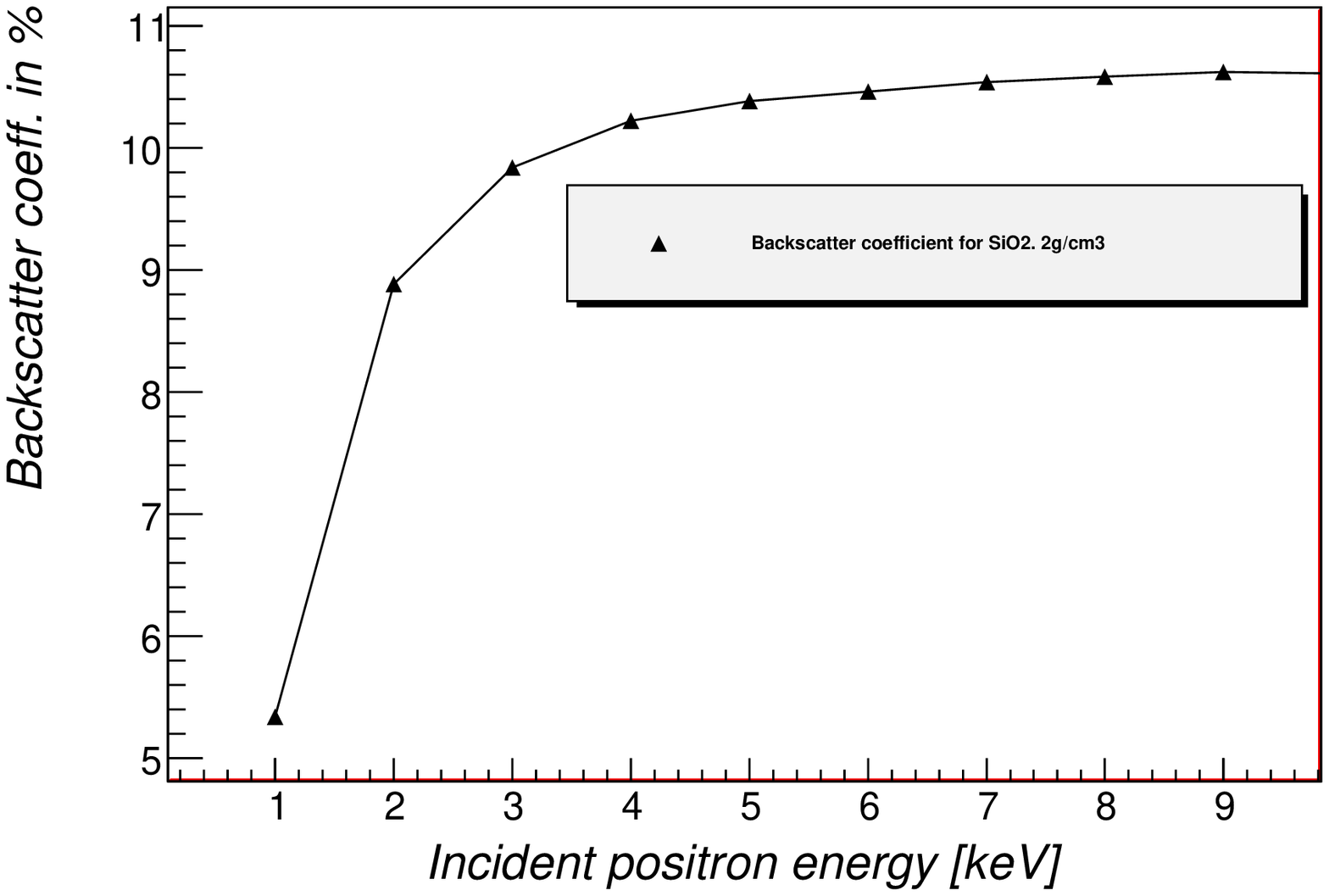}
\caption{\em Backscattering coefficient $\eta_{+}$ for the $SiO_{2}$ target as a function of the positron incident energy.}
\label{backcoeffTarget}
\end{center}
\end{figure}

\begin{figure}[h!]
\begin{center}
%%\hspace{2.cm}\includegraphics[width=0.8\textwidth]{opsdecayrate.eps}
\includegraphics[width=0.8\textwidth]{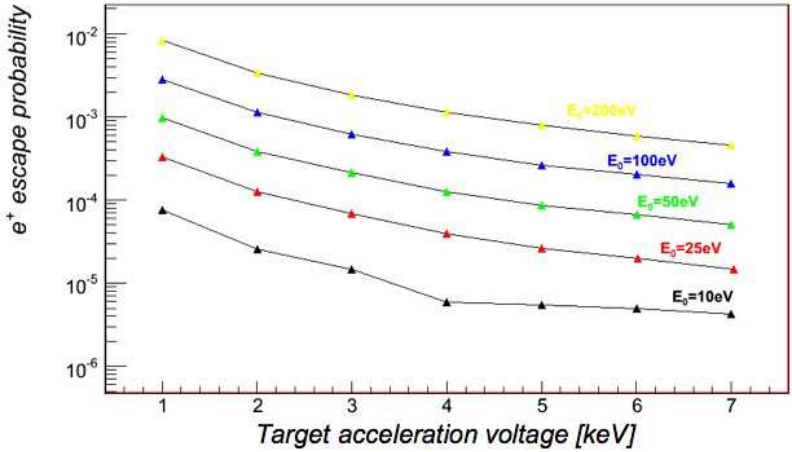}
\caption{\em Escaping probability for the total number of incident positrons at the $SiO_{2}$ target as a function of the implantation voltage for different positron initials energies $E_{0}$.}
\label{BKGLevelBKSCPosSiO2}
\end{center}
\end{figure}

After crossing the carbon foil the positron enters the cavity and it is accelerated to the $SiO_{2}$ target where it may backscatter.
For this estimation, we consider the $SiO_{2}$  target density of $2 g/cm^{2}$. % We first calculate the escaping probability given by the backscattering process and in a second step we include the effect of the transmission of positrons through the carbon foil . 
The backscattering coefficient for SiO$_{2}$ is larger than the one of the 20 nm foil  (Fig. \ref{backcoeffTarget}). In agreement with the results reported in \cite{Mills_transp}, $\eta_{+}$ reaches a constant value of about $10\%$ for positron incident energies larger than 4 keV. The positron escaping probability as a function of the energy is shown in Fig. \ref{BKGLevelBKSCPosSiO2}.  For positron energies between 3-5 keV the escaping probability is between 1$\times 10^{-5}$ and 3$\times 10^{-6}$. These values are larger than the maximum background level that is allowed in order to reach the desired sensitivity. However, this background is clearly overestimated because the energy loss of the positron crossing the carbon foil is not taken into account. An estimation of the mean energy loss \cite{ESTAR} \cite{EGSnrc} gives values of the order of few hundred eV that are significantly larger than the initial positron energy $E_{0}=10$ eV. \\

\begin{figure}[h!]
\begin{center}
%%\hspace{2.cm}\includegraphics[width=0.8\textwidth]{opsdecayrate.eps}
\includegraphics[width=0.6\textwidth]{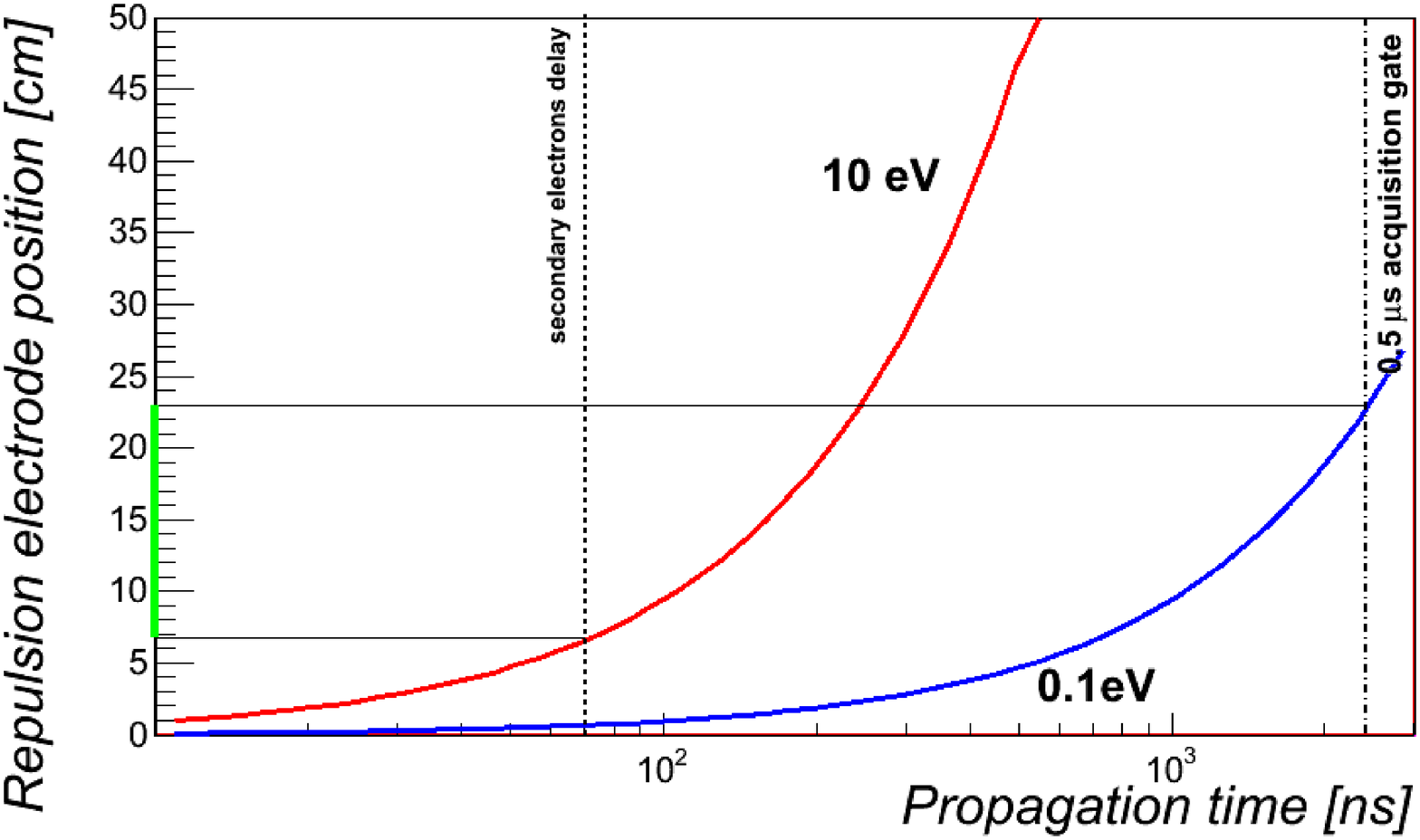}
\includegraphics[width=0.6\textwidth]{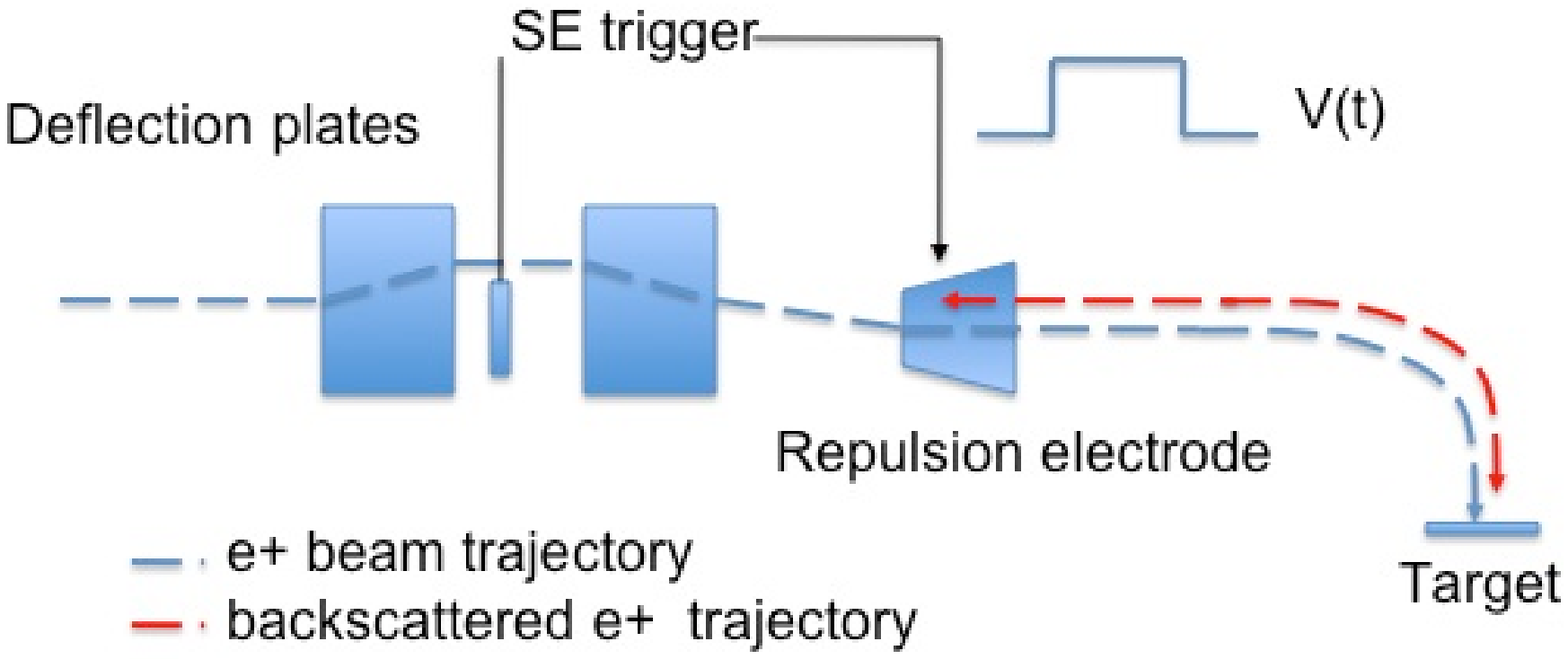}
\caption{\em Scheme of the repulsion electrode used to suppress the backscattered positrons and plot with the optimal position (thick green line on the y-axis). See text for more details.}
\label{repel}
\end{center}
\end{figure}

A possible way to further suppress this background is to redirect to the target the positrons that may backscatter. 
The idea is to exploit the difference in the propagation time between the backscattered positrons (< 10 eV) and the SE used for the trigger (>2 keV). When a SE  triggers the MCP an electrode with a repulsive voltage can be activated to block and redirect back a positron that may have backscattered as shown in Fig. \ref{repel}. The optimal position of the repulsion electrode has been determined assuming that the backscattered positrons have an energy between 0.1 and 10 eV. We defined the minimum distance between the electrode and the target considering the propagation time of the SE from the target to the MCP (70 ns  with d=1 m, E= 2 keV) and the propagation time of the backscattered positron. The maximum electrode-target separation is given by the constraint that the detection probability for a Ps atom that is eventually formed when the positron is re-implanted in the target should still be high enough to guarantee the aimed sensitivity of the experiment. In fact, the propagation time of the positron from the target to the repulsion electrode and back shorten the acquisition time window for such an event. Taking into account the backscattering probability and the Ps lifetime, we estimated that one needs an acquisition time of at least 0.5 $\mu$s. With these requirements the electrode should be positioned in the region indicated in Fig. \ref{repel} with a thick line (green).   

 In this way, the background produced by backscattered positrons from both the carbon foil or the target can be suppressed to values below $< 10^{-7}$.
Table \ref{Bkgtbl_vac}, summarizes the expected background level for the different background sources.

\begin{table}[h!]
  \begin{center}
    {\begin{tabular}{|c|c|c|} \hline
	\multicolumn{2}{|c|}{BACKGROUND} & \\
	\multicolumn{2}{|c|}{SOURCE}	&   expected \\
	\hline
	\hline

	1)&Photon detection loss:\hphantom{00} &  \\
	&Hermiticity\hphantom{00} &  \\
	&Dead Material\hphantom{00} & $<10^{-7}$ \\
	&Resolution\hphantom{00} &    \\
	\hline
	2)&Positron backscattered from &  $<10^{-7}$  \\ 
	&carbon foil\hphantom{00} &   \\
	\hline
	3)& Positron Backscattered from & $<10^{-7}$  \\ 
	&SiO2\hphantom{00} &   \\
	\hline
	4)&Fast \ops\ from &  \\ 
	&carbon foil\hphantom{00} & 5 $\times <10^{-8}$ \\
	\hline
	5)&Fast \ops\ from & $<<10^{-8}$   \\ 
	&target \hphantom{00} &   \\
	\hline
 	\hline
    \end{tabular}}
    \caption{\em Summary of the expected background level for the different background sources.}
  \label{Bkgtbl_vac}
\end{center}
\end{table}

%As it was mentioned before, the operation in bunching mode is essential to enhance the confidence level of a positron hitting the target in order to reach the desired sensitivity. In fact, this suppresses the main source of beam associated background which is expected from electrons and ions due to ionization of the residual gas atoms by positrons and the accidentals with the MCP noise to a level that is estimated to be $9\times 10^{-10}$. 

\section{Sensitivity}\label{sec:sensitivity}
The sensitivity $S_{\rm{o-Ps}\to\rm{invisible}}$ of the experiment is defined as the level at which the first background event is expected:
\begin{equation}
S_{\rm{o-Ps}\to\rm{invisible}} = 1/ ( N_{\ops}\cdot \epsilon_{tot})
\end{equation}
where the terms in the denominator are the integrated number of produced
$\ops$ ($N_{\ops}$) and $\epsilon_{tot}\simeq 1$ is the total efficiency to detect an invisible decay. We neglected the losses in signal efficiency of about 1.5\% arising from the possibility of having 2 or more positrons per bunch. We estimated the rate of these events using $R_{2e^+}= 2 \cdot \tau_{bunch} \cdot R_{e^+}$ where $ \tau_{bunch}=300$ ns and $R_{e^+}= 2.5\times 10^4/s$ is the number of delivered positrons per second on the target in continuous mode. For two or more positrons there is always annihilation energy deposition in the ECAL, hence this effect does not result in a background. 
 The number of \ops /s, $R_{\ops}$, is defined as a product 
\begin{equation}
R_{\ops}=R_{e^+}\cdot \epsilon_{\ops}\cdot \epsilon_{SEtagging}  \cdot \epsilon_{SEcoincidence} \cdot \epsilon_{Bunching}
\end{equation}
where the first factor was defined above, the second one is the efficiency for $\ops$ production (about 30\%) and the third one $\epsilon_{SEtagging}$ is the efficiency of tagging the secondary electrons from the carbon foil (about 20\%). The fourth factor is the efficiency of the coincidence between the SEs from the CF and target  ($\epsilon_{SEcoincidence}=0.04$) and the last one, $\epsilon_{Bunching}=0.1$, are the losses due to the duty cycle of the bunching system. As in our previous search, the length of the gate for the ADCs has to be at least 3 $\mu$s in order to suppress the probability for \ops\ to decay after this time to a level of 10$^{-9}$. 
Therefore, the aimed sensitivity of 10$^{-7}$ can be reached in a 8 hours run ($\approx 1\times 10^7$ observed $\ops$ annihilations). 
For zero signal events observed and no event of background expected, the upper limit at 90$\%$ CL for the branching ratio assuming Poisson statistics is given by:
\begin{equation}\label{eq:BR_opsinv}
Br(\ops \rightarrow invisible) = 2.3/ ( N_{\ops} \cdot \epsilon_{tot}) 
\end{equation}
Solving Eq. (\ref{eq:diffeqs}) with the estimated average number of $\ops$ collisions in the cavity $N_{coll}\simeq 0.9$ (for 5 keV implantation energy of the positrons), shows that a limit on the  $Br(\ops \rightarrow invisible) =1 \times 10^{-7}$ will result in a limit on the photon-mirror photon mixing strength of: 
\begin{equation}\label{limitmixing}
\epsilon \leq  4 \times 10^{-9}.
\end{equation}
This is about one order of magnitude more stringent than the BBN limit (\ref{bbn1}) and can be achieved in a 18 hours run.  
 
Assuming that the DAMA/NaI and DAMA/LIBRA annual signal modulation is generated by elastic scattering of mirror matter, the mixing strength is of the order of $\epsilon \leq  4\times 10^{-9}$, thus a total number of $\simeq$ 93 signal events would be detected in the ECAL during one month of data taking. To be conservative we consider a background level of $1\times 10^{-7}$, thus, about 90 background events are expected which means that a discovery with about 7 $\sigma$ significance could be possible \cite{pdg}.  As explained in Section \ref{sec:setup}, a unique feature of our proposal is the possibility to change the experimental conditions (i.e. the number of the \ops\ collisions with matter), and hence to cross check the results without affecting the background. For an implantation energy of the positron of 3 keV, the number of excess events will be 2 times smaller (46 events) compared to 5 keV positrons.

\section{Summary}\label{sec:conclusion}

In this paper, a proposed search for mirror-type dark matter, looking for \ops\ invisible decays in a vacuum cavity, is presented. In the Introduction,  the mirror matter relevance to the dark matter problem and its link with positronium has been reviewed.
The effect of external fields (electric and magnetic) on the oscillation probability and of the collisions of \ops\ with the cavity walls were estimated. In Section \ref{sec:setup}, the design and the experimental results for the different parts of the experiment have been presented. Section \ref{sec:Bkg_vac} includes the simulation results of the background estimation and Section \ref{sec:sensitivity} the estimated sensitivity. The goal is to reach a sensitivity in the branching ratio of $Br(\ops\to invisible)\simeq 10^{-7}$ to confront the annual modulation signal observed by DAMA/NaI and DAMA/LIBRA (with 8.2$\sigma$ significance)  with mirror dark matter scenarios. In case of a signal observation, the experiment would offer a unique and essential feature, whereby an increase or decrease of the signal rate by a factor $\sim 2$  is possible while keeping the background constant. In case of a null result, this search will provide a limit on the photon-mirror photon mixing strength about one order of magnitude better than presently derived from Big Bang Nucleosynthesis.

\vskip0.5cm

{\bf Acknowledgments}

 We wish to thank Z. Berezhiani, R. Eichler, R. Foot, D. S. Gorbunov, S. Karshenboim, N. V. Krasnikov, C. Lenz-Cesar, V. Matveev,  V. A. Rubakov, and  A. A. Tokareva for the very useful discussions. We acknowledge A. Antognini and F. Kottmann for lending us some carbon foils. 
We are grateful to A. Badertscher, N. A. Golubev, L. Knecht, L. Liszkay, J. P. Peigneux, P. Perez, D. Sillou for their essential help. This work was supported by the ETH Zurich, by the Swiss National Science Foundation, by the Institute for Nuclear Research of the Russian Academy of Sciences and by the CNPq (Brazil). We thank the CERN for its hospitality.

\end{document}